\documentclass[acmlarge]{acmart}

\AtBeginDocument{%
  }

\setcopyright{cc}
\setcctype{by-nc-sa}
\acmJournal{IMWUT}
\acmYear{2025} \acmVolume{9} \acmNumber{1} \acmArticle{210} \acmMonth{3}
\acmDOI{10.1145/3712289}

\acmISBN{978-1-4503-XXXX-X/18/06}

\usepackage{xspace}
\usepackage{enumitem}
\usepackage[capitalise,noabbrev]{cleveref}
\usepackage{tabularx}
\usepackage{listings}
\usepackage{tcolorbox}
\usepackage{tikz}
\usepackage{longtable}
\usepackage{subcaption}
\usepackage{booktabs}
\usepackage{xcolor}
\usepackage{color, colortbl}

\definecolor{Gray}{gray}{0.9}
\definecolor{LightGray}{gray}{0.6}

\usetikzlibrary{backgrounds}

\newcommand{\dataset}{\textit{DiversityOne}\xspace}
\newcommand{\goto}[1]{\textbf{#1}}

\newcommand{\nilogusers}{782\xspace}
\newcommand{\nsensors}{26\xspace}

\newcommand{\warningcell}[1]{}

\newcommand{\vcenteredinclude}[1]{\begingroup
\setbox0=\hbox{\includegraphics[height=1.5em]{#1}}%
\parbox{\wd0}{\box0}\endgroup}

\tcbset{
    mytakeaway/.style={
        colframe=black,           
        colback=white,            
        boxrule=0.4pt,            
        fonttitle=\itshape,       
        sharp corners,            
        coltitle=black,           
        before skip=10pt,         
        after skip=10pt,          
        title={\color{black}},  
        colbacktitle=white,       
        coltitle=black,           
        attach title to upper,    
        boxed title style={
            size=small,           
            boxrule=0.4pt,        
            colframe=black        
        }
    }
}

\usepackage{soul}
\newcommand{\cancel}[1]{}
\newcommand{\change}[1]{#1}

\newcommand{\aau}{Aalborg University\xspace}
\newcommand{\lse}{London School of Economics and Political Science\xspace}
\newcommand{\uc}{Universidad Católica ``Nuestra Señora de la Asunción''\xspace}
\newcommand{\jlu}{Jilin University\xspace}
\newcommand{\ipicyt}{Instituto Potosino de Investigación Científica y Tecnológica\xspace}
\newcommand{\unitn}{University of Trento\xspace}
\newcommand{\amrita}{Amrita Vishwa Vidyapeetham\xspace}
\newcommand{\num}{National University of Mongolia\xspace}

\newcommand{\AAU}{AAU\xspace}
\newcommand{\LSE}{LSE\xspace}
\newcommand{\UC}{UC\xspace}
\newcommand{\JLU}{JLU\xspace}
\newcommand{\IPICYT}{IPICYT\xspace}
\newcommand{\UNITN}{UNITN\xspace}
\newcommand{\AMRITA}{AMRITA\xspace}
\newcommand{\NUM}{NUM\xspace}

\begin{document}

\title[\dataset Dataset]{\textit{DiversityOne}: A Multi-Country Smartphone Sensor Dataset for Everyday Life Behavior Modeling}


\author{Matteo Busso}
\orcid{0000-0002-3788-0203}
\email{matteo.busso@unitn.it}
\authornote{Corresponding author}
\author{Andrea Bontempelli}
\orcid{0000-0001-7037-5797}
%
\author{Leonardo Javier Malcotti}
\orcid{0009-0000-0589-6393}
\affiliation{\institution{University of Trento}\country{Italy}}
%
\author{Lakmal Meegahapola}
\orcid{0000-0002-5275-6585}
\affiliation{\institution{ETH Zurich}\country{Switzerland}}
%
\author{Peter Kun}
\orcid{0000-0003-0778-7662}
\affiliation{\institution{IT University of Copenhagen}\country{Denmark}}
%
\author{Shyam Diwakar}
\orcid{0000-0003-1546-0184}
%
\author{Chaitanya Nutakki}
\orcid{0000-0002-5164-2391}
\affiliation{\institution{Amrita Vishwa Vidyapeetham}\country{India}}
%
\author{Marcelo Dario Rodas Britez}
\orcid{0000-0002-7607-7587}
\affiliation{\institution{University of Trento \& FBK}\country{Italy}}
%
\author{Hao Xu}
\orcid{0000-0001-8474-0767}
%
\author{Donglei Song}
\orcid{0000-0001-6737-6932}
\affiliation{\institution{Jilin University}\country{China}}
%
\author{Salvador Ruiz Correa}
\orcid{0000-0002-2918-6780}
%
\author{Andrea-Rebeca Mendoza-Lara}
\orcid{0000-0003-2301-2891}
\affiliation{\institution{Instituto Potosino de Investigación Científica y Tecnológica}\country{Mexico}}
%
\author{George Gaskell}
\orcid{0000-0001-6135-9496}
%
\author{Sally Stares}
\orcid{0000-0003-4697-0347}
%
\author{Miriam Bidoglia}
\orcid{0000-0002-1583-6551}
\affiliation{\institution{London School of Economics and Political Science}\country{UK}}
%
\author{Amarsanaa Ganbold}
\orcid{0000-0003-4335-6608}
%
\author{Altangerel Chagnaa}
\orcid{0000-0003-2331-3045}
\affiliation{\institution{National University of Mongolia}\country{Mongolia}}
%
\author{Luca Cernuzzi}
\orcid{0000-0001-7803-1067}
%
\author{Alethia Hume}
\orcid{0000-0002-1874-1419}
\affiliation{\institution{Universidad Católica "Nuestra Señora de la Asunción"}\country{Paraguay}}
%
\author{Ronald Chenu-Abente}
\orcid{0000-0002-1121-0287}
%
\author{Roy Alia Asiku}
\orcid{0009-0004-0369-6151}
%
\author{Ivan Kayongo}
\orcid{0009-0007-4429-7335}
\affiliation{\institution{University of Trento}\country{Italy}}
%
\author{Daniel Gatica-Perez}
\orcid{0000-0001-5488-2182}
\affiliation{\institution{Idiap Research Institute \& EPFL}\country{Switzerland}}
%
\author{Amalia de Götzen}
\orcid{0000-0001-7214-5856}
\affiliation{\institution{Aalborg University}\country{Denmark}}
%
\author{Ivano Bison}
\orcid{0000-0002-9645-8627}
\affiliation{\institution{University of Trento}\country{Italy}}
%
\author{Fausto Giunchiglia}
\orcid{0000-0002-5903-6150}
\affiliation{\institution{University of Trento}\country{Italy}}
\email{fausto.giunchiglia@unitn.it}


\renewcommand{\shortauthors}{Busso et al.}


\begin{abstract}
Understanding everyday life behavior \change{of young adults} through personal devices, e.g., smartphones and smartwatches, is key for various applications, from enhancing the user experience in mobile apps to enabling appropriate interventions in digital health apps. Towards this goal, previous studies have relied on datasets combining passive sensor data with human-provided annotations or self-reports. 
However, many existing datasets are limited in scope, often focusing on specific countries primarily in the Global North, involving a small number of participants, or using a limited range of pre-processed sensors. These limitations restrict the ability to capture cross-country variations of human behavior, including the possibility of studying model generalization, and robustness. To address this gap, we introduce \textit{DiversityOne}, a dataset which spans eight countries (China, Denmark, India, Italy, Mexico, Mongolia, Paraguay, and the United Kingdom) and includes data from \nilogusers college students over four weeks. DiversityOne contains data from \change{\nsensors smartphone sensor modalities} and 350K+ self-reports. As of today, it is one of the largest and most diverse publicly available datasets, while featuring extensive demographic and psychosocial survey data. DiversityOne opens the possibility of studying important research problems in ubiquitous computing, particularly in domain adaptation and generalization across countries, all research areas so far largely underexplored because of the lack of adequate datasets. 

\vspace{1em}
\tikzstyle{background rectangle}=[thin,draw=black]
\begin{center}
\begin{tikzpicture}[show background rectangle]
\node[align=center, text width=40em, inner sep=1em]{
\textbf{Dataset Catalog:} \url{https://livepeople.disi.unitn.it} \\
\textbf{\dataset Website:} \url{https://datascientia.disi.unitn.it/projects/diversityone/}
};
\node[xshift=3ex, yshift=-0.7ex, overlay, fill=white, draw=white, above 
right] at (current bounding box.north west) {
\textbf{Important Links}
};
\end{tikzpicture}
\end{center}

\end{abstract}



\received{20 February 2007}
\received[revised]{12 March 2009}
\received[accepted]{5 June 2009}

\maketitle

\section{Introduction} \label{sec:intro}
Human behavior, routines, habits, and social practices are deeply interwoven with the events, places, interactions, and technologies that compose our everyday lives. Each of these elements contributes to larger behavioral patterns influenced by our continuous engagement with pervasive devices. Our emotional experiences fluctuate daily, shaped by factors such as sleep quality \cite{khalid2024sleepnet}, which, in turn, impact our physical activity, forming both recurring weekly patterns \cite{tseng2016assessing} and longer-term habits \cite{harari202019}. Everyday actions, such as eating and drinking, also have significant effects on our health \cite{biel2018bites, santani2018drinksense}, with additional factors like interactions with technology playing important roles in everyday life  \cite{das2019multisensor}.

Smartphones, as prevalent devices in daily life, influence our behavior, often in nuanced ways that can be both positive \cite{lin2021revisiting} and negative \cite{2017-SOCINFO}. Their use is linked to our internal states \cite{elhai2018depression, buda2021outliers}, and understanding how people engage with their smartphones could provide valuable insights into human behavior. Prior research has shown that smartphone sensors can infer a range of behavioral aspects in young adults, including mood \cite{servia2017mobile, meegahapola2023generalization}, depression \cite{xu2023globem}, activities \cite{haresamudram2021contrastive, assi2023complex}, social context \cite{mader2024learning, hernandez2024proximity}, and eating or drinking episodes \cite{bae2017detecting, santani2018drinksense, thomaz2016automatic, biel2018bites}. This makes smartphone sensor data valuable for improving app design, supporting user well-being, and creating more effective interventions. However, there are profound variations in daily behaviors across different cultural and geographic backgrounds \cite{foner2020introduction}\footnote{\change{We operationalize the concept of culture through the theory of social practices outlined in \cref{subsubsec:soa-socio}. This concept recognizes that cultures are complex and multidimensional, often coexisting or overlapping within the same geographic areas \cite{yuval2004gender}. Additionally, it acknowledges that certain regularities emerge from specific socio-cultural contexts or from the communities of practice that develop locally. The use of terms like ``country,” ``culture,” and ``geographic region” throughout the paper reflects this distinction.}}. Ignoring these variations has often hindered model generalization \cite{meegahapola2023generalization, khwaja2019modeling}, limiting the effectiveness of behavior inference models when applied in diverse real-world settings.

Despite the richness of smartphone sensor data, a critical challenge persists: the lack of datasets that capture behavioral diversity across different cultural and geographic contexts. Existing datasets and research often draw from specific populations typically in the Global North \cite{khwaja2019modeling, meegahapola2023generalization}, and may not fully account for the cultural, environmental, and social norms that shape daily routines and smartphone usage patterns in other regions \cite{phan2022mobile}. This limits the development of machine learning models capable of generalizing across different populations. For instance, eating habits \cite{tobin2018dinner}, sleep routines \cite{stacker2023sleep, cheung2021considering}, and social interactions \cite{gsir2014social, parady2021comparative} vary widely between countries, affecting how individuals interact with their smartphones and how other daily life aspects are inferred from sensor data \cite{lopez2017self, phan2022mobile}. Consequently, models trained on data from one country or region may perform poorly when applied to another. This underscores a fundamental challenge in generalizing machine learning models to real-world applications \cite{meegahapola2023generalization, assi2023complex}. Addressing this challenge requires comprehensive datasets that not only capture a variety of sensor data but also integrate cultural, social, and geographical diversity. Such datasets would enable researchers to study behavioral differences and similarities across populations, to test model robustness in diverse contexts beyond typical experimental settings\change{, and to investigate how to combine model generalization and adaptation to local data to capture local and individual behaviors.} 

The \dataset dataset, developed as part of the large-scale European project \textit{``WeNet - The Internet of US''}\footnote{\url{https://doi.org/10.3030/823783}.} \cite{2025WenetPaper},
provides rich, cross-country data for studying everyday life behavior through smartphone sensors and self-reports. \dataset spans data from college students in eight countries: 
{China} (\jlu, \textbf{\JLU}),
{Denmark} (\aau, \textbf{\AAU}), 
{India} (\amrita, \textbf{\AMRITA}), 
{Italy} (\unitn, \textbf{\UNITN}), 
{Mexico} (\ipicyt, \textbf{\IPICYT}), 
{Mongolia} (\num, \textbf{\NUM}), 
{Paraguay} (\uc, \textbf{\UC}), and 
the {United Kingdom} (\lse, \textbf{\LSE}). 
The study involved more than 18,000 students, of which \nilogusers agreed to participate in an intensive longitudinal survey of four weeks. Based on the innovative \texttt{iLog app}~\cite{2014-PERCOM}, adopted for the data collection, \dataset includes raw data from \nsensors smartphone sensors, such as accelerometers, gyroscopes, and GPS, as well as derived information like notification interactions, app usage, activities, and step counts, during these four weeks. These data streams are organized into six category bundles (connectivity, environment, motion, position, app usage, and device usage), allowing modular downloads for ease of use. Additionally, participants provided multiple in-situ self-reports each day, through time diaries, detailing their activities, locations, social contexts, and moods, together with daily reports on sleep quality and daily expectations. This combination of self-reported data and sensor data provides a comprehensive view of behavior, making \dataset a valuable resource for studying the impact of smartphone technology across various contexts in both the Global North and the Global South \cite{biel2018bites, khalid2024sleepnet, qin2019cross, grammenos2018you}. \change{Additionally, DiversityOne enables more computational social science-oriented studies focused on human behavior (see, e.g., \cite{2017-SOCINFO, zhang2021putting, bontempelli2020learning}).}

The \dataset dataset supports the development of machine learning models capable of inferring user behaviors, with potential applications in domain generalization, domain adaptation\change{, transfer learning}, and self-supervised learning. It is a valuable research resource in ubiquitous computing, human-computer interaction, and machine learning, offering a broader scale than previously available datasets. Furthermore, \dataset adheres to the European General Data Protection Regulation (GDPR) \cite{GDPR2016}, with ethical approvals from all participating institutions. The study design and data collection processes were developed by an interdisciplinary team of computer scientists, social scientists, interaction designers, ethicists, and legal experts to meet the highest standards. This extensive dataset has been used in a few initial publications by different authors, showing its various uses \cite{meegahapola2023generalization,  kammoun2023understanding, assi2023complex,meegahapola2024m3bat, mader2024learning,girardini2023adaptation,mercado2023social}, and demonstrating its potential for reuse in studies aiming to understand behaviors across countries through smartphone sensing. The paper describes three contributions:

\begin{itemize}[wide, labelwidth=!, labelindent=0pt]

\item[\textbf{Contribution 1:}] We publicly release \dataset, one of the largest and most geographically diverse datasets combining questionnaires from 18K+ participants and passive smartphone sensor data and self-reports from \nilogusers participants across eight countries, covering both the Global North and the Global South. The dataset includes raw sensor data from \nsensors modalities, providing researchers with a flexible resource for analyzing and developing machine learning models across multiple domains, such as behavior recognition, cross-cultural studies, and multimodal time-series modeling. This dataset facilitates explorations of topics like model generalization and domain adaptation across countries, with detailed self-reports on demographic and psychosocial variables, adding depth for nuanced, multifaceted analyses that can be conducted independently or integrated with sensor data.

\item[\textbf{Contribution 2:}] We provide an in-depth description of the study design and data collection, which was tailored to enable effective data gathering across multiple countries, each with unique languages, privacy regulations, and cultural contexts (Section~\ref{sec:method}). To manage the diversity and sensitivity of the collected data, the dataset was consolidated under rigorous privacy and ethical standards, hosted in a GDPR-compliant secure environment, and respecting the ownership rights of all contributors.

\item[\textbf{Contribution 3:}] We synthesize key insights, lessons learned, and recommendations derived from some initial studies leveraging \dataset, highlighting unique opportunities for further studies in smartphone sensing, behavioral modeling, and machine learning. We provide the insights under two broader themes---Lessons learned from (i) study design and cross-country data collection (\cref{subsec:lessons_design}) and (ii) multi-country sensor data analysis (\cref{subsec:lessons_analysis}). Our 12 recommendations, which we see as actionable, aim to provide ideas for future research, enhancing the adaptability and accuracy of smartphone-based behavioral studies in diverse global contexts.

\end{itemize}

The remainder of the paper is organized as follows: \cref{sec:relatedworks} provides background and context for the methodology used to create \dataset, reviewing existing datasets for modeling everyday behaviors through smartphone sensors alongside various forms of ground truth. \cref{sec:method} details the methodology and protocol adopted to collect diverse, cross-country data, with specific attention to variations in cultural contexts of student's daily routines and data privacy regulations. \cref{sec:validation} presents key results to validate the data collection approach and the dataset's quality. \cref{sec:availability} outlines the resources available for accessing and using the dataset, highlighting compliance with privacy and copyright standards. \cref{sec:discussion} explores potential use cases for the dataset, provides insights and recommendations, emphasizing its value in studying behavioral patterns across different regions and advancing machine learning applications. Lastly, \cref{sec:conclusion} summarizes the paper’s contributions and impacts.

\section{Background and Related Work}\label{sec:relatedworks}
\subsection{Smartphone Sensing for Behavior Modeling}

Smartphone sensing data is widely used to train machine learning models to infer user behaviors in real-world environments \cite{meegahapola2020smartphone}. Developing these models involves continuously collecting data from sensors like accelerometers, gyroscopes, GPS, Bluetooth, and app usage logs to create labeled datasets. These datasets serve as ground truth for various behaviors, including daily activities, social interactions, mental states, sleep, and dietary habits \cite{meegahapola2020smartphone}. These labeled datasets fuel machine learning models, from classic approaches (e.g., random forest, decision tree, support vector machines) to deep learning techniques, enabling the detection of behavioral patterns and predictions on new, unseen data. Once trained, these models are deployed in-the-wild, running on users' smartphones to provide real-time insights into daily routines, mental well-being, and contextual changes, all without requiring active input from the user. Traditionally, such models have relied heavily on continuous sensing modalities—such as movement patterns and physical proximity—to infer user behavior. More recently, however, interaction sensing modalities (e.g., app usage, typing events, and notification interactions) have gained significance, providing rich insights into users’ engagement patterns and internal states.

\subsection{Continuous and Interaction Sensing Modalities} Smartphone sensors can be broadly classified into two types \cite{meegahapola2020smartphone}: continuous sensing and interaction sensing, both essential for understanding nuances of everyday life.

Continuous sensing involves passive data collection without direct user interaction. For instance, accelerometers and gyroscopes detect movement patterns (e.g., walking, running, or inactivity) that offer insights into physical activity levels and types, relevant for understanding behavioral patterns, energy expenditure, and emotional states, such as depression or anxiety indicated by prolonged inactivity or erratic movements~\cite{elhai2018depression, canzian2015trajectories}. Step count, derived from inertial sensors, further enhances activity monitoring, revealing daily routines and physical well-being. Additionally, proximity sensors and Bluetooth signals can indicate social contexts by identifying close proximity to others, which helps assess social isolation or engagement~\cite{meegahapola2020smartphone, lane2010survey}. GPS and WiFi data are indispensable for determining semantic locations, enabling insights into whether users are at home, at work, or engaged in recreational activities—each of which correlates with mental states or other behavioral patterns~\cite{santani2018drinksense, meegahapola2023generalization, servia2017mobile}.

Interaction sensing captures smartphone user engagement, providing insights into attention, productivity, and emotional states. App usage, for example, can infer attention spans, productivity, and even stress levels, as excessive or reduced engagement with certain apps may be linked to anxiety or depressive symptoms~\cite{guracho2023smartphone}. Studies have shown that interactions with social media, in particular, are tied to mental well-being~\cite{buda2021outliers}. Similarly, frequent typing events, rapid touch interactions, or ignoring notifications can signal heightened stress or distraction \cite{vahedi2018association}. Notifications, app-opening frequencies, and related metrics also provide important context for attention spans and can identify potentially compulsive behaviors, such as frequent app-checking, which may indicate underlying mental health concerns \cite{meegahapola2020smartphone, carlo2019numbers}.

Combining continuous and interaction sensing modalities offers a comprehensive view of behavior, allowing machine learning models to capture nuanced and expressive aspects with a high level of detail. Currently available public smartphone sensing datasets, however, often lack this level of depth and diversity in sensing modalities, limiting the ability to fully understand human behavior at scale. Even if many modalities are present, most publicly available datasets only offer pre-processed versions of data, limiting the use of such datasets for diverse purposes. Therefore, there is an unmet need for raw smartphone sensor datasets that integrate continuous and interaction sensing modalities, offering richer insights into behavioral patterns. 

\subsection{Effect of Country Diversity on Sensing and Self-reported Ground Truth}

Together, continuous and interaction-sensing modalities build a holistic picture of user behavior, capturing both passive and longitudinal patterns and immediate interactions that correlate with mental and emotional states. However, sensor data and self-reported behaviors used as ground truth for training machine learning models can vary significantly across countries, influenced by cultural, social, and environmental differences \cite{phan2022mobile}.

For example, sensor data like accelerometer readings or step counts may reflect distinct physical activity patterns based on country-specific factors, such as urban infrastructure, transportation habits, and climate \cite{ICLEI_UrbanTransport, ZeroHourClimate_UrbanPlanning, EPA_ClimateTransportation}. In countries where walking or cycling is common, step counts may show higher physical activity levels, while regions reliant on driving will have lower activity levels. Similarly, proximity sensor and Bluetooth data capture different social interaction patterns, reflecting cultural norms around personal space, social gatherings, and work environments \cite{ozella2021using, janssen2024tracking, sekara2014strength, hernandez2024proximity}. In collectivist cultures, for instance, users may exhibit more frequent close proximity with others, while individualistic cultures may show more solitary sensor readings \cite{trumbull2001bridging, triandis2001individualism}. Cultural attitudes towards technology could also shape app usage and notification interactions \cite{bombardi2017exploring}. Social media may dominate app engagement in some countries, whereas others might have more work-focused or communication-restricted technology use \cite{poushter2018social, cheng2021prevalence}. Beyond sensing modalities, the way behaviors are reported as ground truth---such as moods, stress, context, or activity types---varies culturally \cite{mesquita1992cultural, meegahapola2023generalization, sebe2005multimodal}. Cultural norms can influence self-reported stress or mood, leading to different labels for similar sensor data patterns across countries \cite{schmidt2019wearable, mesquita1992cultural, meegahapola2024m3bat}. In machine learning terms, such cross-country variations introduce challenges related to data covariate shift and label shift, both of which complicate generalization \cite{bickel2009discriminative, koh2021wilds}.

Despite these evident variations, most prior studies overlook country-level differences, relying on homogeneous datasets from specific regions, often within the Global North \cite{meegahapola2020smartphone, phan2022mobile}. This oversight leads to models that struggle to generalize effectively to diverse populations and cultural contexts. In some instances, even though cross-country data were available, analysis has not focused on country differences \cite{servia2017mobile}. Without datasets that capture country-specific data and behavioral patterns, machine learning models are likely to produce incorrect inferences when applied outside the regions in which they were trained. To improve model robustness, globally diverse datasets are essential for exploring behavioral differences across populations, allowing models to account for the variations in behaviors and sensor data across countries and cultures. Addressing this gap is one of the main objectives of the \dataset dataset.

Furthermore, collecting large-scale, passive smartphone sensing data with accurate labels is challenging due to significant costs, time, and effort, especially when collecting data from diverse international participants \cite{yfantidou2023beyond}. Datasets in this field typically involve fewer than 100 participants (see Table~\ref{tab:relatedwork}) since managing continuous data streams, conducting longitudinal data collection, and providing accurate behavioral labels make scaling these studies complex. Scaling up to engage hundreds of participants requires intensive recruitment, sustained user engagement, and technical infrastructure to handle continuous sensor data collection over extended periods. This is partly why the field of ubiquitous computing (ubicomp) lags behind fields like computer vision or natural language processing, where collecting millions of labeled images, videos, or text is relatively more straightforward and cost-effective, enabling faster progress. However, data collection across countries presents additional challenges, as cultural, behavioral, and infrastructure differences influence how people use smartphones, complicating the labeling process. Unlike images, videos, or text, which can be labeled through standardized methods or crowdsourcing, passive sensing data requires detailed, context-specific labeling of behaviors like mood, activity, and social interaction, often relying on time-consuming self-reports. Cross-country collection further complicates matters with logistical issues such as language differences, varying ethical standards, and compliance with different data privacy laws. These challenges make large-scale, diverse data collection costly and time-intensive. The \dataset dataset’s achievement in collecting data from \nilogusers participants demonstrates the commitment and resources required to overcome these hurdles.

\change{\subsection{Sociological Theory and Approaches to Study Design} 
}\subsubsection{Cultural Diversity as Social Practices} \label{subsubsec:soa-socio}
\change{The concept of culture has been extensively examined in social sciences. Previously, culture was viewed as a cohesive framework that influenced attitudes and practices through socialization \cite{swidler1986culture}. However, recent studies suggest that culture is fragmented and diverse among social groups \cite{lizardo2016dual}. This shift in perspective sees culture as complex structures of quasi-rules that individuals can use strategically \cite{bourdieu1990logic, sewell1992theory} and highlights the need to analyze relationships among various cultural influences, linking them to distinct phenomena and calling for a more nuanced psychological understanding \cite{cerulo2010mining}. }

The study design of the \dataset dataset focuses on a particular definition of culture, deriving from the social practice theory \cite{wittgenstein1953philosophical,goffman1975asylums,giddens1979central,giddens1984society,bourdieu1977outline,bourdieu1990logic,Dreyfus1991world,schatzki2001practice,reckwitz2002toward}. This theory frames human behavior as composed of daily practices that produce societal outcomes and influence individual skills and mindsets, contributing to social structure and culture. According to \citet{shove2012dynamics}, social practices consist of three key components:

\begin{itemize}
\item \textbf{Material}: The physical objects or resources, such as a car or a membership, that enable the execution of a particular practice.
\item \textbf{Competence}: The knowledge, skills, and abilities that make a certain practice possible.
\item \textbf{Meaning}: The cultural and symbolic elements that give significance to social practices, motivating individuals to perform them in alignment with societal norms.
\end{itemize}
\noindent
For instance, environmental consciousness may motivate individuals to choose public transportation, aligning with behaviors like cycling, waste sorting, or adopting a vegetarian lifestyle. These motivations, shaped by material access (such as bike availability or recycling facilities), personal competence, and meaningful commitment to sustainability, collectively define behaviors that, when widely adopted within a community, become recognized social practices \cite{shove2005consumers,ropke2009theories}. 

\change{Thus, there is a distinction between the practitioner (individual) and the social practice (community). While interconnected, social practices exist independently and are established at the social level. Indeed, \citet{ropke2009theories} noted that practices consist of recognizable, interconnected elements that individuals reproduce, with new members being constantly recruited. In this sense, individuals are ``carriers of practices'' recruited based on their backgrounds, not merely choosing practices by utility \cite{reckwitz2002toward}, and the distribution of practices often reflects social inequality, with varying cultural perceptions of what constitutes ethical practice. Participation in a practice leaves lasting effects, such as knowledge and skills, which facilitate future involvement in that practice in a path-dependent process~\cite{ropke2009theories}. Finally, diversity is socially recognized, and practices are inherently social, resembling one another across different contexts \cite{reckwitz2002toward}. In other words, social practices exhibit regularity - models of how certain daily practices are typically and habitually performed in (a considerable part) of a society \cite{holtz2014generating}.}

\change{This perspective enhances our understanding of human behavior in real-world environments, where the activities recorded by sensors may be limited in terms of quantity, quality, and context representation. Identifying whether an individual belongs to a community of practice and their ``career'' within it enables the inference of important characteristics, such as the nature of their activities—distinguishing, for example, between an athlete's training session or recreational activity—the tools utilized, such as appropriate clothing and supportive technologies, and the likelihood of recurring behavioral patterns over time. This understanding is particularly valuable in situations where data may be limited or less accurate.}

\change{Furthermore, gaining insight into the context in which these practices occur allows for culturally informed comparisons. This encompasses both the individual perspective, which considers the factors that may facilitate or impede engagement in a practice or access to a community, and the social perspective, which examines how the reference community typically interacts with that practice.}

\change{\subsubsection{Capturing Social Practices}} \label{subsubsec:soa-meth}
Researchers utilize various methodologies to effectively capture behaviors and lifestyles over time. In European countries, for example, the Harmonized European Time Use Surveys (HETUS) measure time spent on various activities. At the same time, the Experience Sampling Methodology (ESM) focuses on the interoceptive, or internal, aspects of behavior \cite{csikszentmihalyi2014validity, myin2022esm}. Time-use diaries, commonly used in HETUS, are intensive longitudinal surveys where participants self-report activity sequences over a day, detailing the frequency and duration of their actions to reveal intricate social patterns. Participants typically complete these diaries at regular intervals, tracking each activity, which enables a detailed, objective view of daily routines and interactions \cite{sorokin1939time}. In contrast, ESM prompts participants to report their thoughts and behaviors over days or even months, detailing their personal points of view on their daily experiences. A comprehensive review by \citet{van2017experience} highlights the strengths of ESM in capturing these aspects.

\change{Our approach combines these two perspectives, namely time-use diaries and ESM with smartphone sensor data. For every sensor pattern that indicates an action or habit—such as accelerometer data, Movement Activity Labels, or GPS positions—researchers can reconstruct the individual's social and personal context. 
A study review explores this combination of continuous sensing and interaction sensing with self-report data for the well-being of young adults~\cite{meegahapola2020smartphone}.
Here, self-report data can serve as ground truth events. Conversely, in scenarios like interactive classification in the wild \cite{bontempelli2020learning}, algorithms such as skeptical learning can use sensor information to validate user annotation. Furthermore, this approach enables the researcher to observe human behavior from the individual point of view within the community of practice, specifically her understanding and interpretation of the context in which the activity occurs (see, e.g., \cite{zhang2021putting}).}

\subsection{Currently Available Public Smartphone Sensing Datasets}

\begin{table}[btp]
\caption{\label{tab:relatedwork} Public available datasets for activity and context recognition in the wild. Duration is expressed in days. The number of pilot sites indicates whether the countries are in the Global North (N) and/or Global South (S). (+) Sensors are divided between smartphones and (+) other devices. No '+' means only smartphone sensors. (*) The GLOBEM data collection was done during a semester for 10 weeks each year, for 4 different years.}
\begin{tabularx}{\textwidth}{lrrcXr}
\toprule
\textbf{Datasets}%
& \multicolumn{3}{c}{\textbf{Coverage}}%
& \multicolumn{2}{c}{\textbf{Purpose}}\\  
\cmidrule(rl){2-4}\cmidrule(rl){5-6}
 & \multicolumn{1}{c}{\textbf{Sample}}         & \multicolumn{1}{c}{\textbf{Days}} & \multicolumn{1}{c}{\textbf{Sites}} & \multicolumn{1}{c}{\textbf{Self-reports}}                          & \multicolumn{1}{c}{\textbf{Sensors}}   \\
\midrule
MDC (2013)   \cite{laurila2013}               & 185                      & 365                      & 1 N                 & Location   
                                        & 12+14                     \\
StudentLife  (2014)  \cite{wang2014studentlife}          & 48                      & 70                      & 1 N                 & Sleep-related                                 & 10                    \\
ExtraSensory  (2017)  \cite{vaizman2017recognizing}      & 60                      & 7                       & 1 N                & Activity                                      & 8+2                 \\
Real-life HAR  (2020)  \cite{gonzalez2020}               & 19                      & 28                      & 1 N                &  .                                            & 4                     \\
ContextLabeler  (2021)  \cite{campana2021contextlabeler} & 3                       & 14                      & 1 N                 & Activity                                      & 18                    \\
Qwantify  (2022)   \cite{wilson2022qwantify}          & 242                     & 7   & 1 N                & Desire, emotion,    well-being                & . \\
LifeSnaps  (2022) \cite{yfantidou2022lifesnaps}           & 71                      & 120 & 4 N & Location, Mood, Step Goal & 0+23 \\
ETRI  (2022) \cite{chung2022real}                         & 22                      & 28                      & 1 N               & Activity, Location,   Relation, Mood          & 10+4                \\
SmartUnitn2  (2018)  \cite{li2022representing}           & 158                     & 28                      & 1 N               & Activity, Location,   Relation, Mood          & 28                    \\
LAUREATE   (2023) \cite{laporte2023laureate}             & 42                      & 91                      & 1 N                 & Activities, Health                            & 0+6                   \\
GLOBEM  (2023) \cite{xu2023globem}                        & 534                     & 280*                     & 4 N              & Standard scales (personality, physical, mental and social well-being)                             & $\sim$10              \\
EgoADL  (2024)  \cite{sun2024multimodal}                 & 30                      & 5                       & 1 N                & Automated labels                              & 4+2                   \\

 & & & & & \\
\textbf{\dataset}                                               & \textbf{\nilogusers}            & \textbf{28}             & \textbf{8 (3N,5S)}        & \textbf{Activity, Location,   Relation, Mood} & \textbf{26}   \\        
\bottomrule
\end{tabularx}
\end{table}

Several datasets that leverage smartphone and smartwatch sensors for activity and context recognition have been developed across various research fields. \cref{tab:relatedwork} compares key publicly available datasets, highlighting their sample size, duration, data collection sites, self-reported questionnaires or annotations via intensive longitudinal survey, and number of sensors used, both collected via smartphone and other devices. While these datasets contribute valuable insights, most are constrained by small sample sizes and short data collection durations, which limit their ability to capture daily routines and behavioral patterns. For instance, although datasets like MDC~\cite{laurila2013}, StudentLife~\cite{wang2014studentlife}, and Real-life HAR~\cite{gonzalez2020} extend beyond short-term studies, they still lack the diversity required to observe routine behaviors that often necessitate more extended observation periods and cross-country differences.

Datasets that focus solely on continuous sensor data, such as Real-life HAR \cite{gonzalez2020}, are limited in capturing the complexity of human activities because they do not include annotations reflecting users' experiences or contexts. Moreover, a growing trend involves using experience sampling methodologies (ESM), where users frequently self-report their experiences, to capture behaviors and contextual nuances. Datasets like the Qwantify app~\cite{wilson2022qwantify} provide valuable insights into emotions, desires, and well-being. Still, sensor data in these collections often remain secondary, serving mainly as context rather than central to the analysis. Hence, \cancel{we believe it is fair to say that}the most valuable datasets combine sensor data with self-reports about the collected data. Such datasets can be further divided into those with in-situ (reports provided about the current moment) and retrospective (reports provided about past periods or days) self-reports~\cite{meegahapola2020smartphone}. A notable example of retrospective self-reporting is the work of \citet{krumm2013placer}, which uses time diaries (i.e., similar to the HETUS approach described in \cref{subsubsec:soa-meth}) for labeling. However, the lack of direct user interaction during labeling can lead to inaccuracies and contextual inconsistencies. In contrast, datasets like StudentLife~\cite{wang2014studentlife}, ExtraSensory~\cite{vaizman2017recognizing}, and ContextLabeler~\cite{campana2021contextlabeler} rely on in-situ self-reports from users. Still, they are often limited to specific activities and do not adhere to a standardized reference, which introduces social and cognitive biases.

GLOBEM \cite{xu2023globem} presents a large-scale dataset that enables cross-dataset generalization analysis, which is particularly valuable for examining behavioral patterns across different periods and university environments. Its longitudinal nature, spanning over \change{four years (ten weeks of data collection each year)}, allows researchers to assess how models perform across diverse academic settings over time. However, the dataset predominantly focuses on the USA, limiting its applicability for cross-country studies. Hence, while GLOBEM provides essential insights, its geographic concentration contrasts with the more globally diverse scope of \dataset, which includes participants from eight countries across the global north and south, offering a broader foundation for generalization and cross-country behavior modeling. Moreover, \dataset provides more fine-grained and raw sensor data and frequent in-situ self-reports, giving researchers much flexibility and depth in their analysis. 

Among existing datasets, LifeSnaps \cite{yfantidou2022lifesnaps} is the most comparable to \dataset in terms of countries of data collection. However, despite the richness, the dataset falls short in crucial areas such as sample size, diversity of data collection countries, and breadth of sensor coverage. LifeSnaps focuses primarily on four European countries and only has a sample size of 71 participants, whereas \dataset spans both the global north and south, including a broader cultural and geographic range, with \nilogusers participants. This comprehensive scope makes \dataset a unique resource for understanding everyday life behavior across diverse countries, surpassing other available datasets in terms of depth and scale.

\subsection{The \dataset Dataset}
\change{Our dataset aims at making a valuable contribution by differing significantly from previous works in several key aspects, as outlined below. First, as described in \cref{subsubsec:soa-meth}, our dataset methodology follows best practices from earlier studies by measuring behaviors and psychosocial traits using standardized and validated methods. We propose a combination of self-reported annotations and sensor data from smartphones. This integration enables a nuanced understanding of social practices, considering routine behaviors alongside contextual data about both physical actions and mental states. While many current datasets aim to capture such elements, few achieve the same depth of blending multiple methodologies with sensor data. This multifaceted approach allows for a more accurate, holistic view of human behavior, providing fine-grained labels essential for training robust machine learning models that reflect the complexity of everyday life. Second, to the best of our knowledge, there are currently no publicly available smartphone datasets specifically tailored to examining social practices (see \cref{subsubsec:soa-socio}). This innovative framework allows researchers to enrich their understanding of individuals by situating them within a broader community of profiles that share similar competencies, materials, and meanings. For instance, it distinguishes between those engaged in professional activities and those pursuing recreational interests, highlighting each group's distinct personalities and values. This nuanced approach encourages researchers to explore the diversity of daily activities in a way that emphasizes the cultural context of behaviors, thereby creating a more comprehensive understanding of how social practices shape individual actions and interactions.}

\change{Finally, our sample size and the breadth of data collected internationally, with a particular emphasis on regions in the global south —underrepresented in prior research— is rather unique. In addition to our theoretical and methodological approach, these characteristics enable a more comprehensive exploration of human behavior, shedding light on the diversity among people and their similarities. In summary, by integrating diverse cultural and socio-demographic factors, \dataset facilitates a deeper exploration of behavior modeling, personalization, and cross-cultural adaptation at a more nuanced level. Moreover, incorporating these factors fosters cross-cultural and multidisciplinary analysis of human behavior rather than promoting it solely as a machine learning benchmark dataset~\cite{orr2024ai,raji2021ai}.}

\section{Methodology}\label{sec:method}

The design and collection of this dataset posed the challenge of ensuring consistent data comparability between different countries while mapping social practices in a culturally sensitive and diversity-aware manner. Behaviors and practices vary considerably according to time of day and location and are shaped by different cultural and social factors. Local cultural variations also impact ethical and privacy considerations, influencing the adoption of data collection protocols, legal frameworks, and public perceptions. Furthermore, the diversity of participating organizations and the resources available in each context significantly shaped the data collection process.

\begin{figure}[!hbt]
    \includegraphics[width=\textwidth]{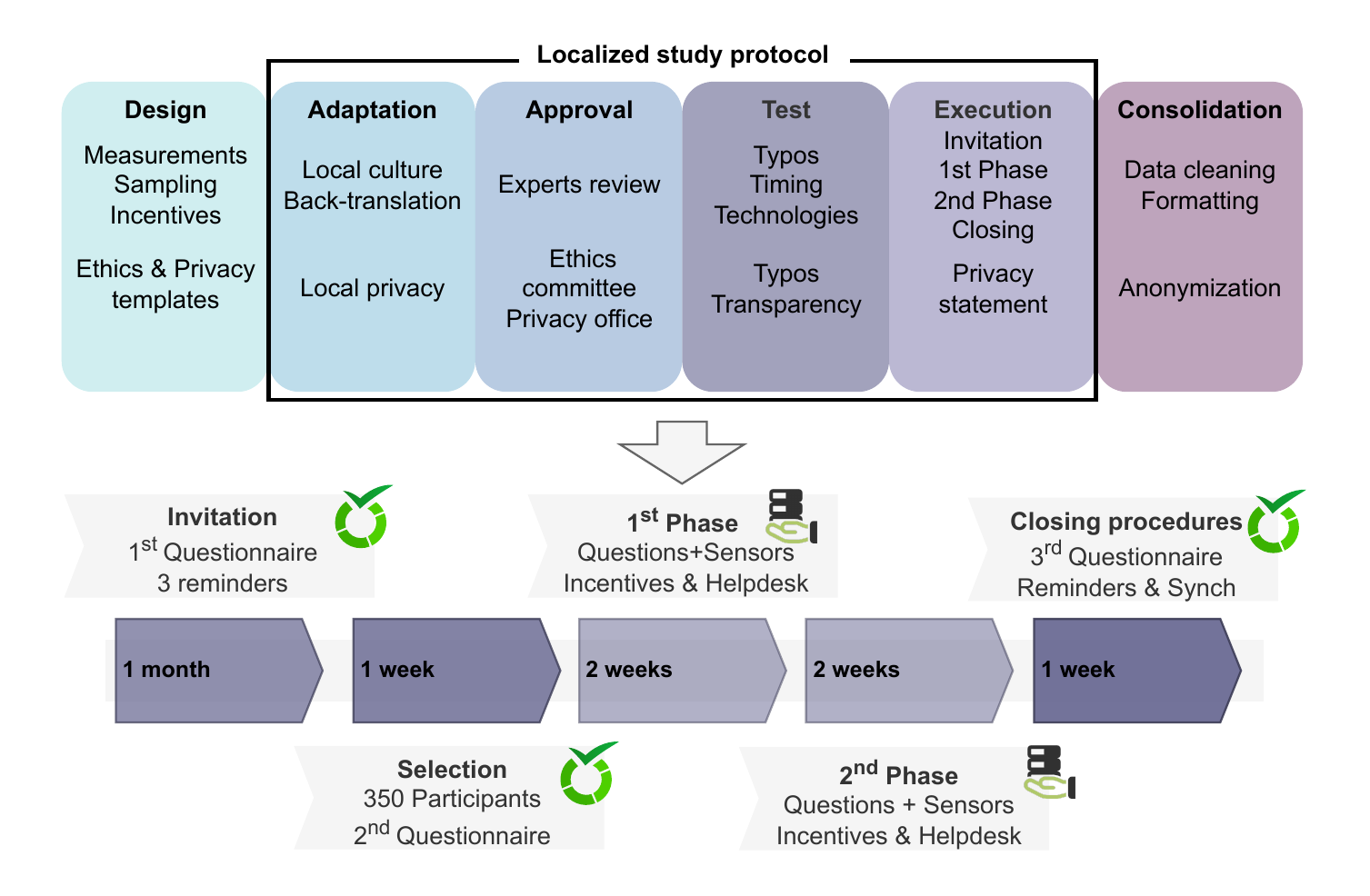}
    \Description[Overview of the study set-up and data collection]{The upper part shows the sequence of steps of the study protocol: design, adaptation, approval, test, execution, and consolidation. The lower part shows the data collection timeline: 1 month for the participant invitation and first questionnaire, 1 week for participant selection and second questionnaire, 2 weeks for the first phase of the data collection, 2 weeks for the second data collection, and 1 week for the third questionnaire and closing procedures.}
    \caption{Study set-up and data collection process. Invitation, Selection, and Closing procedures were done using the LimeSurvey platform. The 1st Phase and 2nd Phase were done using the iLog mobile app.}
    \label{fig:ExpPhasesDetailed}
\end{figure}
\noindent
\cref{fig:ExpPhasesDetailed} shows the study set-up and the survey protocol, as applied consistently across all sites, with minor adjustments which still maintained comparability (refer to \cref{subsec:protocol}). The protocol was developed with an adaptive approach which allowed to address local perspectives and requirements. A team of experts, including computer scientists, social scientists, interaction designers, incentive designers, ethicists, and legal representatives, developed the study design, survey protocol, and procedures to ensure compliance with ethics and privacy standards. The experiment partners then adapted these materials in a way to align with local contexts. Each adapted version was validated by the initial team or an authorized body to ensure functionality, adherence to standards, and comparability.

For the study design, \UNITN and \LSE identified the need to measure human behavior over time by adopting both questionnaires and an intensive longitudinal survey approach \cite{iida2023using, mcneish2021measurement} enabled by an adapted version of the iLog app \cite{2014-PERCOM}\footnote{iLog is an app designed to collect intensive longitudinal survey data (in the form of daily self-reports or hourly time diary self-reports) along with passive smartphone sensor data. The initial version of the app, made available by Zeni et al. \cite{2020-zeni1,2014-PERCOM}, was modified with new features to facilitate the collection of the \dataset dataset.}. The content of the questions and the technology were tailored to address local needs (refer to \cref{subsec:questionnaire,subsec:ils}). The questions from the iLog app and the questionnaires were subsequently translated into multiple languages, as detailed in \cref{subsec:questionnaire}. \change{Furthermore, this type of intensive longitudinal survey often faces challenges related to participant dropout, which can impact the quality of the results. According to~\cite{de2014dropouts}, dropout in longitudinal studies can originate from \textit{(i)} failure to locate the research unit, \textit{(ii)} failure to contact the potential respondent, and \textit{(iii)} failure to obtain cooperation.}

\change{Moreover, repeated assessments can become burdensome, potentially leading to decreased compliance over time. However, according to \citet{wrzus2023ecological}, factors such as the total number of assessments or the frequency of assessments do not impact compliance. Indeed, the burden on respondents can be mitigated, as participants who agree to participate are already willing to adhere to demanding schedules. There is a consensus that offering monetary incentives can enhance participation rates. \citet{wrzus2023ecological} also discusses how it is possible to hypothesize that including break options could improve compliance, although their meta-analysis did not find evidence to support this idea. }

\change{A survey protocol was developed in response to these considerations. This protocol involved sending invitations on the same day to all students enrolled at various universities via their institutional email accounts, addressing the issues of locating research units, and contacting potential respondents. We tackled the challenge of obtaining cooperation by ensuring the participants received all the necessary information in advance and incentivizing participation in the one-month survey with monetary compensation. Adjustments were made to the invitation language, sample selection, and incentive amounts to better align with local requirements, as detailed in \cref{subsec:sample}.}

Data collection involved handling personal data across both European and non-European countries. European legislation mandates that any processing of data related to European citizens or conducted using tools based in Europe must comply with the General Data Protection Regulation (GDPR). Consequently, \UNITN prepared procedures and documentation to ensure GDPR compliance. These processes were then adapted for different contexts to ensure effective data management, clear communication with participants, and adherence to privacy requirements where necessary (refer to \cref{subsec:privacy}).

The following sections outline the strategy for adapting measurement tools, privacy protocols, and methods to collect high-quality data for studying diversity in human behavior and social practices over time. Each section addresses both the general approach and specific local adaptations.

\subsection{Questionnaires about Demographics and Social Practices} \label{subsec:questionnaire}

The administered questionnaires measured various aspects of social practices and identified communities of practice relevant to students. Given the wide range of social practices, we focused on a specific set, including social relationships (both online and offline), cultural and sports activities, daily commuting, and shopping and cooking habits. The questionnaire items were designed to capture elements of competence, materiality, and meaning associated with these practices (see \cref{sec:relatedworks}). Each question and scale thus provide foundational information on components of social practices. For example, a profile of an individual's shopping and cooking habits can be built by examining the types of food purchased (materials), cooking skills (competencies, e.g., buying convenience foods \textit{vs.} cooking daily), and motivational values (meaning, e.g., hedonism \textit{vs.} universalism). This combination of components yields a nuanced measure of specific social practices, allowing cross-country comparisons to identify diverse practices. Due to the extensive number of questions, we divided the questionnaire into three parts and administered it throughout the data collection, as mentioned in \cref{fig:ExpPhasesDetailed}.

\begin{enumerate}
    \item The \textbf{First Questionnaire} was sent via an invitation email to all students at each university participating in the study. This questionnaire covered demographic information, cultural interests, and leisure activities, as well as questions about online and offline social interactions. It also included standard scales for measuring personality traits using the \textbf{Big Five Inventory}~\cite{donnellan2006mini} and attitudes toward values measured through the \textbf{Basic Values Survey}~\cite{gouveia2014functional}.
    \item The \textbf{Second Questionnaire} was administered only to participants involved in the sensor and survey data collection. It primarily explored specific social practices such as daily commuting, cooking, grocery shopping, and physical activities. Additionally, a second set of personality questions based on the \textbf{Jungian Scale for Personality Types} \cite{jung1971psychological,briggs1995gifts,mascarenas2016jungian,wilde2011jung} was included, along with a second set of values questions based on the \textbf{Human Values Survey} \cite{schwartz1994there,schwartz2001extending}.
    \item The \textbf{Third Questionnaire} was administered to the same participants as the second questionnaire. It gathered feedback on the user's experience with the data collection app and assessed a \textbf{Multiple Intelligence Scale}~\cite{tirri2008identification}.
\end{enumerate}
\noindent
The structure of the questions reflects the goal of measuring different components of social practices. We adopted standard scales to assess aspects related to meaning—often intangible and challenging to quantify—while additional questions aimed at understanding the other two components of social practices, namely competencies and materials. These components were evaluated through questions about specific practices, such as commuting, sports, cultural activities, and cooking and shopping habits.
To administer the questionnaires, we used \textit{LimeSurvey}~\cite{LS}, a tool widely used in social sciences to protect user privacy. The full questionnaires,  translations and adaptations are available in the data catalog, see \cref{sec:availability}.

\subsubsection{Questionnaire Translation, Testing, and Adaptation}
Each experiment site received the English version of the questionnaire. This version was later translated into the local language and adapted to the local context. We used a validated translated questionnaire whenever available for the standard scales mentioned above. Then, in case of absence, translations were completed by an expert translator and validated through panel and back-translation processes \cite{cha2007translation} by independent translators.
Adaptations were made to tailor specific questions and answer options to each site. This was necessary for questions such as field of study, which varies depending on the educational programs offered by each institution, and accommodation, which differs based on personal choices, available facilities, and cultural norms. For example, at \NUM, there is substantial migration to the capital for work and study, intertwining with students' educational journeys and influencing their choices. At \JLU, all students live on the university campus, contrasting with other institutions that offer various accommodation options, including university-provided, public, or private housing. Further adaptations addressed unique local needs. At \JLU, response options related to nationality were modified to exclude countries whose sovereignty is not recognized by the local government. Additionally, \JLU adapted the list of social media and applications list to include those popular in both Western and Asian markets.

Certain items related to sexuality (“To have sexual relationships; to obtain sexual pleasure”) and religiosity (“To believe in God as the savior of humanity; to fulfill the will of God”) from the Basic Values Survey~\cite{gouveia2014functional} were also modified. The content was deemed inappropriate in some locations, such as \AMRITA and \UC, and alternative options like “love” were suggested. For the religiosity item, an option with other religious and spiritual figures, such as Buddha, was explicitly provided for \AMRITA and \JLU. Finally, aware of the cross-cultural variation of personality traits in the case of \JLU, the locally developed personality trait scale ~\cite{zhang2019development} was adopted in addition to the Western version, thus facilitating a comparison.

\noindent
After translation and adaptation, the questionnaires were tested locally. They were distributed to approximately 30 participants, including project partners and university students. This allowed to validate the translations and assess completion times.

\subsection{Intensive Longitudinal Surveys and Passive Sensor Data Collection} \label{subsec:ils}

\begin{figure}[t]
    \includegraphics[width=\textwidth]{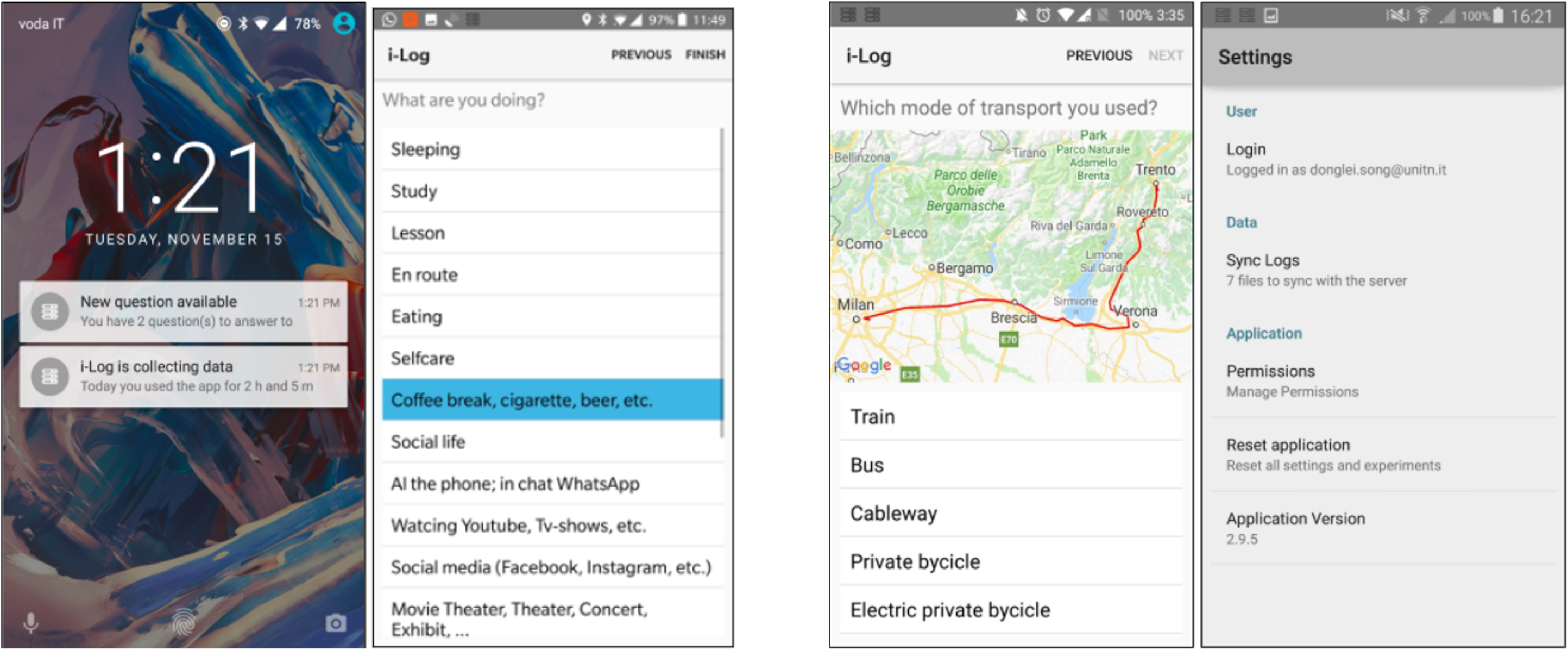}
    \Description[iLog app screenshoots]{Sequence of screenshots of iLog application showing the notification when the application is running in the background, two types of questions and settings.}
    \caption{The iLog app adopted for intensive longitudinal survey and sensor data collection.}
    \label{fig:ilogapp}
\end{figure}

Temporal aspects are essential for understanding how everyday life behavior evolves over time. Intensive longitudinal surveys are valuable for observing human behavior as it changes, especially when paired with detailed sensor data from everyday devices. Thus, we adopted the \texttt{iLog}~\cite{2014-PERCOM}, an App that effectively collects this information through user interactions and data from smartphone sensors\footnote{The use of iLog and collected data in various experiments has been described in \cite{2020-zeni1,2017-SOCINFO,KD-2017-PERCOM,2017-ICSC,2018-PERCOM2}. At the time of this study, only Android devices were supported. Currently, the \change{new app version} runs on both Android and iOS devices.}. 
\change{\cref{fig:ilogapp} shows the application screenshots.} The app’s ability to gather user self-reports and passive sensor data makes it unique (see \cite{runyan2013smartphone,wang2014studentlife,kreuter2020collecting} for other similar tools). This dual capability significantly enhances traditional time diary methods \cite{2018-PERCOM1,sorokin1939time}, mainly when structured \cite{hellgren2014extracting} to enable real-time user responses. In addition, \change{the application enters power-saving mode when the smartphone reaches around 8\% of the phone battery charge}, thus limiting cases of non-response due to the smartphone being turned off.

\subsubsection{Diary Methods and Daily Routines}\label{sec:td}
Daily diaries, as a form of self-reports, are foundational for studying daily human behavior. Following standards set by HETUS and ATUS, three types of daily diaries were implemented\footnote{Answer options were supplemented with explanatory notes, not included here for readability. Complete documentation is available in the catalog (see \cref{sec:availability}).}:

\begin{itemize}
\item \textbf{Morning and Evening Diaries}: The first diary collects information at the start and end of each day. At 08:00 AM, participants received two qualitative questions about sleep quality and daily expectations, listed in \cref{tab:morning_questions}. At 10:00 PM, participants responded to questions in \cref{tab:evening_questions} regarding \textit{(a)} a rating of their day, \textit{(b)} any challenges encountered, \textit{(c)} how they addressed these challenges, and \textit{(d)} reflections on the COVID-19 pandemic.
\item \textbf{Time Diaries}: The second time diary follows the HETUS standard (\cref{tab:td_main}), covering main activities and moods\footnote{The questions design aligns with the definition proposed in \cite{li2022representing, giunchiglia2022context} concerning a model of the situational context of a person which can be applied in organizing and reasoning about massive streams of sensor data and annotations. This alignment expands the dataset’s potential, enabling its adoption in fields such as knowledge graph modeling and engineering.}. Throughout the study, participants were asked to answer four brief questions every thirty minutes during the first two weeks and every hour during the last two weeks:
\begin{itemize}
     \item \textit{What are you doing?}: Participants select from 34 activities, such as sleeping, eating, and working.
     \item \textit{Where are you?}: The current location is identified from 26 categories, including home, workplace, university, and restaurant.
     \item \textit{Who is with you?}: Participants indicate their company, with options like ``alone,'' ``with my partner,'' and ``with friends.''
     \item \textit{What is your mood?}: Participants rate their mood on a five-point scale ranging from happy to sad.
\end{itemize}
If a participant reports engaging in ``eating,'' ``traveling,'' or ``doing sport,'' they are prompted to answer additional questions (\cref{tab:td_sub}):

\begin{itemize}
    \item For eating, participants select foods and drinks from 20 categories, such as rice, potatoes, meat, and beer (adapted from~\cite{gatica19}).
    \item For sports, participants specify the type of sport from nine options, such as jogging, running, or water sports.
    \item For traveling, participants state \textit{(a)} the purpose, selecting from seven categories (e.g., study, social life), and \textit{(b)} the mode of transport from 16 options (e.g., car, bus).
\end{itemize}
\item \textbf{Snacks Diary}: As listed in \cref{tab:tb_snack}, these questions document food and drink consumption outside of main meal times. These prompts appear every two hours, allowing participants to select multiple items.
\end{itemize}
\noindent
To reduce the participant burden \cite{ROLSTAD20111101, eisele2022effects}, which can lead to non-response missing data or wrong answers \cite{bison2024impacts}, notification breaks were incorporated (see Table \ref{tab:break}). These breaks allow participants to pre-fill answers or pause notifications temporarily. For example, participants can select the “Go to sleep” option in the evening to silence notifications for six hours or pause notifications for two hours during classes or sports.
Moreover, each notification expired only after 12 hours, allowing participants to choose not to respond immediately. They could accumulate a maximum of 24 unanswered notifications. After this limit, the application deleted the oldest notifications and marked them as ``Expired'' in the dataset.

\begin{table}[]
    \footnotesize
    \centering
    \caption{Time diaries collecting contextual information about the participant every half hour.}
    \label{tab:td_main}
    \begin{tabularx}{0.97\textwidth}{p{8cm}X}
    \toprule
    \textbf{A3. What are you doing?}&
    \textbf{A4. Where are you?}\\
    \midrule
    \begin{enumerate}[leftmargin=*]
        \item Sleeping
        \item Personal care
        \item Eating \goto{(go to A3c,  \cref{tab:td_sub})}
        \item Cooking, Food preparation \& management
        \item Study/work group
        \item Lecture, seminar, conference, university meeting
        \item Did not do anything special
        \item Rest/nap
        \item Break
        \item Walking
        \item Travelling \goto{(go to A3a1, A3a2, \cref{tab:td_sub})}
        \item Social life
        \item Happy Hour, Drinking, Party
        \item Phone/Video calling
        \item In chat on Internet or reading, sending e-mail
        \item Surfed or seeking, reading information via Internet
        \item Social media (Facebook, Instagram, etc.)
        \item Watching TV, video, YouTube, etc.
        \item Listening to music
        \item Reading a book, periodicals, news, etc.
        \item Movie Theatre Concert\dots
        \item Entertainment Exhibit, and Culture
        \item Others Entertainment and Culture
        \item Arts
        \item Hobbies
        \item Games
        \item Freetime study
        \item Sport \goto{(go to A3b, \cref{tab:td_sub})}
        \item Voluntary work and participatory activities
        \item Household and family care
        \item Grocery Shopping
        \item Other Shopping
        \item Work
        \item Other
    \end{enumerate}&
    \begin{enumerate}[leftmargin=*]
        \item Home apartment, room
        \item Home garden, patio, courtyard
        \item Relatives Home
        \item House (friends others)
        \item Classroom, Laboratory
        \item Classroom, Study hall
        \item University Library
        \item Other university places
        \item Canteen
        \item Other Library
        \item Gym, swimming pool, Sports centre\ldots
        \item Grocery Shop
        \item Supermarket\ldots
        \item Street markets
        \item Shops, shopping centers, indoor markets, other shops
        \item Café, pub, bar
        \item Restaurant, pizzeria, Street food vendor
        \item Movie Theatre Museum\ldots
        \item In the street
        \item Public Park/Garden
        \item Countryside, mountain, hill, beach
        \item Workplace, office
        \item Weekend home or holiday apartment
        \item Hotel, guesthouse, camping site
        \item Another indoor place
        \item Another outdoor place
    \end{enumerate} \\
    \toprule
    \textbf{A5. With whom are you?}&\textbf{A6a. What is your mood?}\\
    \midrule
    \begin{enumerate}[leftmargin=*]
        \item Alone
        \item Friend(s)
        \item Relative(s)
        \item Classmate(s)
        \item Roommate(s)
        \item Colleague(s)
        \item Partner
        \item Other
    \end{enumerate}&
    \begin{enumerate}[leftmargin=*]
    \item \vcenteredinclude{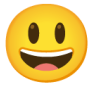}
    \item \vcenteredinclude{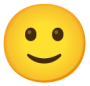}
    \item \vcenteredinclude{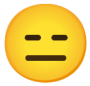}
    \item \vcenteredinclude{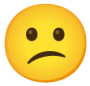}
    \item \vcenteredinclude{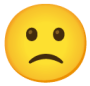}
    \end{enumerate}
    \\
    \bottomrule
\end{tabularx}
\end{table}

\subsubsection{Sensor Data} 
During the study, iLog collected data from all available smartphone sensors, spanning continuous and interaction sensing modalities (see \cref{sec:relatedworks}), at \change{a very high level of granularity} to capture the breadth of daily behaviors, which can be studied through various analytical techniques. Appendix~\ref{app2:sensors} \change{details the sensors and their collection frequency or whether they are event-based}. The collected sensors support research in the following areas:
\begin{itemize}
\item \textbf{Connectivity}: includes Bluetooth Low-Energy, Bluetooth Normal, cellular network, connected WiFi network, and discovered WiFi Networks, measuring connections to nearby devices.
\item \textbf{Environment}: includes ambient temperature, light, pressure, and relative humidity sensors, which are hardware-based and available only on compatible devices.
\item \textbf{Motion}: captures participants' movements in different contexts and includes accelerometer, activities, gravity, gyroscope, step counter, and step detector.
\item \textbf{Position}: includes location, magnetic field, proximity sensors, tracking physical positioning and interaction with the device.
\item \textbf{App Usage}: tracks social media and app interactions, including headset plug, music playback, notifications, and running applications.
\item \textbf{Device Usage}: tracks device interaction patterns, i.e., airplane mode, battery charge, battery level, doze, ring mode, screen status, touches, and user presence sensors.
\end{itemize}

\subsubsection{Intensive Longitudinal Survey Translation, Testing, and Adaptation}
Like the previous questionnaires conducted with LimeSurvey, the intensive longitudinal self-report questions and responses were translated into local languages. In addition to the local coordinators, approximately 15 participants downloaded and used the iLog app for two weeks to assess user experience and technical stability in a separate test. We performed adaptations to accommodate specific local interests and needs. Some partners were interested in investigating eating behaviors and eating disorders. Thus, additional questions were added to cover food consumption and dietary routines, balanced with standard time diary items, to minimize invasiveness. Regional variations in response options were also included to reflect local food and beverage consumption. Additionally, \IPICYT opted to collect the notification at a half-hour frequency for the entire month.

Further technical adaptations were made to the iLog app code provided by \cite{2020-zeni1}. For instance, iLog typically relies on Google services to receive notifications from the server. However, these services were unavailable in certain regions, such as at \JLU, or in areas with limited cellular network coverage. An ``offline'' version of iLog was developed to overcome this issue. Here, the notification schedules are downloaded at the beginning using WiFi connections and displayed on participants’ smartphones at planned times, removing the dependency on Google services or data connections. Participants can then synchronize their collected data via WiFi as needed. Initially designed for \JLU, the offline version was also implemented at \AMRITA and \IPICYT for their data collections.

\subsection{Target Population and Sampling Strategy} \label{subsec:sample}

We focused on a specific target population to ensure effective comparisons across study sites: \textit{college students}. Globally, students tend to have similar routines, such as attending classes, studying, working part-time, and participating in extracurricular activities. This shared lifestyle makes it easier to compare social practices, revealing commonalities and differences across diverse contexts—i.e., the ``unity in diversity" of their behaviors.

The sampling strategy was designed with specific criteria to represent the student population. Except for \JLU and \IPICYT, all registered students at each university were invited to participate in the study and complete the first questionnaire. From the participant pool, individuals with incompatible smartphones (e.g., Android OS below 5.0) or low survey attendance were excluded. The sample was designed to be balanced by gender, age, and academic department to minimize routine variation due to scheduling, community characteristics, and demographics. Thus, the dataset is comparable to existing studies, such as \cite{wang2014studentlife}, and aligns with similar studies that follow this methodology and collect comparable variables \cite{li2022representing, 2017-SOCINFO}.

\subsubsection{Recruitment Strategy}
Students at each pilot site received an invitation to their institutional email accounts, followed by four monthly reminders. Recruitment strategies differed at \JLU and \IPICYT, employing snowball sampling via WeChat and WhatsApp. After this initial phase, 350 students were selected according to predefined sampling criteria to ensure a minimum of 250 active participants, accounting for typical longitudinal survey dropout rates of 30\% to 70\% \cite{gustavson2012attrition}. In the pilot sites where achieving this threshold proved challenging, all students who completed the questionnaire and owned compatible smartphones were invited to participate.

The selected students received a second email with attachments, including instructions for downloading the iLog app, survey information, and a privacy statement detailing data processing methods. This email also included the second questionnaire. Participants were given approximately five days to download the app and complete the registration, receiving additional official communications halfway through the survey (along with the third questionnaire) and at the study’s conclusion. The helpdesk provided daily personalized support. \cref{tab:participants} summarizes the sampling and recruitment strategy outcomes.

\subsubsection{Recruitment Adaptations}

Specific adaptations were made to accommodate local needs. During the invitation phase, some modifications were implemented at \JLU and \IPICYT. At \JLU, the survey invitation was posted on social channels and in various WeChat groups, reaching about 5,000 students. Reminders were shared on these channels rather than being sent directly to participants. At \IPICYT, recruitment was conducted through direct participant engagement, with reminders delivered verbally and via messages in a WhatsApp group created specifically for the study. Additionally, at \IPICYT, the selection phase occurred before the first questionnaire was distributed.

\subsection{Incentive Strategy}\label{sec:incentives}

A group of experts designed the incentive strategy. It was informed by sociological literature, particularly the work of \citet{singer2013use}, who extensively reviewed types of incentives (monetary and non-monetary) and their effectiveness in reducing respondent bias and dropout rates. The review suggests that monetary incentives are generally more effective than gifts, and prepaid incentives outperform random prizes. Based on these findings, an incentive strategy was developed, focusing primarily on fixed payments.
\cref{tab:incentives} reports the incentive strategy adopted in the different pilots. While no incentives were adopted for the invitation questionnaire, in the first and second phases of iLog data collection, selected students received monetary rewards for completing at least 85\% of notifications. Additionally, daily prizes and three final prizes were awarded in both phases. 

\begin{table}[t]
    \centering
    \caption{\label{tab:incentives} Incentives were paid upon completing at least 85\% of the notifications after the first and second two-week periods. Additionally, daily and final prizes were awarded through random selection. Each pilot site adapted the incentive protocol based on local practices from similar studies they had conducted previously.}
    \begin{tabular}{lccccc}
    \toprule
     & \multicolumn{2}{c}{\textbf{Payments}} & \textbf{Daily} & \multicolumn{2}{c}{\textbf{Final Prizes}}\\
     \cmidrule(rl){2-3}\cmidrule(rl){5-6}
        & \textbf{1st Phase} & \textbf{2nd Phase} & \textbf{Prizes} & \textbf{1st Phase} & \textbf{2nd Phase} \\
    \midrule
    \JLU    & 100 rmb    & 100 rmb   & 1 of 20 rmb   & 3 of 88 rmb   & 3 of 88 rmb \\
    \AAU    & 150 kr     & 150 kr    & 5 of 40 kr    & 3 of 800 kr   & 3 of 1200 kr \\
    \AMRITA & - & - & - & - & - \\
    \UNITN  & 20 €       & 20 €      & 5 of 5 €      & 3 of 100 €    & 3 of 150 € \\
    \IPICYT & - & - & - & - & - \\
    \NUM    & 10k MNT    & 10k MNT   & 5k MNT        & 100k MNT      & 150k MNT \\
    \UC     & 25k GS     & 25k GS    & 10 vouchers   & 1 restaurant voucher & 1 restaurant voucher \\
    \LSE    & -          & -         & -             & £150 (1/50)   & £150 (1/50) \\
    \bottomrule
\end{tabular}
\end{table}

\subsubsection{Incentives Adaptation}
The incentives were adjusted according to the cost of goods in each country. In addition, they were tailored to local cultures and circumstances. Thus, in the case of the \LSE, it was decided to offer a prize of £150 drawn randomly for every 50 participants, aligning with the incentive methods proposed at the same institute. Similarly, at \AMRITA, no monetary incentives were implemented; instead, participation certificates were provided, reflecting the local culture of engagement in science. This strategy was particularly successful for questionnaires but did not yield the same results for the intensive longitudinal survey, as discussed in \cref{sec:limits}. This lack of success may also be attributed to the novelty of the approach. Lastly, at \IPICYT, no monetary incentives were used either. Instead, there was a focus on social engagement by creating a WhatsApp group, which facilitated multiple meetings and exchanges of interests and opinions between participants and researchers. 

\subsection{Ethics and Privacy}\label{subsec:privacy}

Addressing privacy in cross-country research requires adapting to diverse regulations, social norms, and individual attitudes toward privacy \cite{acquisti2015privacy, capurro2005privacy, ess2005lost}. Privacy management requires careful consideration of ethical and legal aspects tailored to each context. Consequently, each university's principal investigator or legal deputy was designated as data controller, thus covering a legal and ethical role of responsibility. Specific data processing responsibilities were also delegated to \UNITN, which managed data collection and management, as outlined in \cref{subsec:management}.

From an ethical perspective, the data controller in each country appointed a local ethics committee to support and approve project activities. If universities or research centers lacked an ethics committee, the local institution designated a suitable one. To facilitate this process, \UNITN, in collaboration with its ethics committee and privacy office, developed templates and documentation that were then adapted to meet local requirements.

Given that the iLog app operates within Europe, the methodology adopted a Eurocentric approach, benefiting from the comprehensive framework provided by the GDPR, which governs personal data processing. \UNITN supplied all necessary documentation—such as privacy statements, information protocols, and additional guidelines—beginning with GDPR standards and extending them with clauses as required by local regulations. In essence, regulations in each country that do not conflict with the GDPR were considered complementary.

This approach ensured that each country's implementation adhered to procedures and documentation based on European legislation while accommodating local laws. For example, in the case of \IPICYT, any participant request for data deletion must be fulfilled within three days—a timeframe stricter than GDPR requirements and incorporated into the local documentation and protocols.

\subsection{Data Collection}\label{subsec:protocol}
The academic calendar and various aspects of university life greatly influence students' routines, often more so than other factors. To ensure the ecological validity of the study, data collection at all sites was carried out during the academic semester. This approach guarantees that participants likely adhered to similar academic routines as their peers in other countries.
The questionnaires, longitudinal time diaries, and passive sensor data were collected consistently across the eight pilot sites. The data collection process spanned approximately ten weeks, divided into the following phases (see \cref{fig:ExpPhasesDetailed}):

\begin{itemize}    
    \item \textbf{Invitation}: The first phase involved sending an email with a survey description, an invitation to complete the first questionnaire, and information on the second part of the survey. This invitation was followed by three weekly reminders for students who had not yet completed the survey. 
    \item \textbf{Selection}: At this stage, a subset of eligible participants was selected for the second part of the survey. Selection criteria included consent to personal data processing and possession of a compatible Android smartphone.
    \item \textbf{First Phase}: This phase began with distributing the second questionnaire to the selected participants, followed by a reminder after one week. Along with the second questionnaire, participants received an email with instructions on downloading the iLog app, along with a brief user manual.
    \item \textbf{Second Phase}: During this final data collection phase, the third and final questionnaire was sent, followed by a reminder after one week. In this phase, the frequency of iLog time diary prompts was reduced.
    \item \textbf{Closing Procedures}: At the end of the survey, participants received a final email with instructions on app uninstallation and a last reminder, where necessary, to complete the second and third questionnaires.
\end{itemize}
\noindent
Daily reports were generated throughout the data collection to monitor iLog survey responses and quickly identify potential issues. Based on these reports, local field supervisors contacted inactive participants every three days to provide support as needed. Participants were also contacted with the results of the daily prize draws (see \cref{sec:incentives}). Moreover, due to timing issues, primarily related to technical challenges at \JLU and delays in ethics committee approvals at \AMRITA and \IPICYT, data collection in these countries began several months later than initially planned. 

\subsection{Data Management} \label{subsec:management}

As discussed in \cref{subsec:privacy}, ethical and legal considerations led to establishing protocols to ensure data ownership and privacy compliance, including copyright management, data cleaning, and anonymization, described below.

\subsubsection{Data Ownership and Copyright}

At the end of each data collection, \UNITN, acting as the data processor, securely transferred the entire raw dataset—including personal data—to the designated local data controllers. \UNITN maintained a copy of the non-anonymous data for further preparation, thus sharing a processed copy of the cleaned dataset with all data controllers and researchers involved in refining and analyzing the data. Towards the end of the project, each data controller licensed their cleaned and anonymized dataset to \UNITN, establishing shared management and distribution policies, partly described in \cref{sec:availability}. This procedure ensured local data controllers retained ownership while facilitating legally compliant data sharing for research purposes.

\subsubsection{Data Cleaning and Anonymization}

To ensure GDPR compliance and achieve high data quality, the dataset underwent \textit{Data Cleaning} and \textit{Anonymization}. Data cleaning included standardizing the dataset and addressing non-random missing data due to technical errors. Variable and value labels were adapted for clarity, making the dataset understandable to third-party researchers. On the other hand, Anonymization aimed to prevent individual identification through (i) Personal Data Anonymization, (ii) Network Anonymization, and (iii) GPS Anonymization. 

Regarding personal data anonymization, all personal information, such as email addresses, home addresses, names, and surnames, was removed from the three dataset types (online questionnaire, time diaries, and sensor data). Unique identifiers were assigned to ensure consistency across these datasets. Considering Network Anonymization, potential re-identification sources, such as the WiFi network connected to the smartphone, cellular networks, and available WiFi networks, were anonymized. Relevant sens r file columns were anonymized using irreversible hash functions applied to WiFi network names. Regarding GPS Anonymization, high-precision GPS data were intentionally truncated to maintain anonymity while preserving scientific usability. Anonymization included identifying points of interest (POIs) where participants spent significant time and providing generalized location data. Two versions of this dataset, named \textit{RoundDown} and \textit{POI}, were created. To further protect privacy, a single research institution can access only one of these datasets to minimize re-identification risks associated with combining the two versions.

It is important to note that, despite these measures, the dataset could still allow re-identification if cross-referenced with other datasets. For this reason, the dataset is not publicly available online and can only be accessed under specific conditions (see \cref{sec:availability}).

\section{Validation} \label{sec:validation}
Our dataset provides a rich and diverse array of information, making it possible to extract valuable insights into students' behavior across different pilot sites. Below, we present key descriptive statistics\footnote{\change{The statistics are compatible with those reported in previous studies analyzing the dataset, as they applied data filtering.}} and focus on crucial dimensions that highlight study participation and the diversity observed in the students' everyday lives.

\subsection{Overall Study Engagement}

\begin{table}[tb]
    \centering
    \caption{\label{tab:participants} \change{Number of participants for each of the three questionnaires and data collection. ``iLog signed'' and ``iLog data'' columns show the total number of students who logged into the iLog app and those who actively contributed with data, respectively. Additionally, demographic information of the participants involved in the iLog data collection is reported.}}
    \begin{tabular}{llrrrrrrrr}
    \toprule
    \textbf{Country} &
    \textbf{Acronym} &
    \textbf{1st QU} &
    \textbf{2nd QU} &
    \textbf{3rd QU} &
    \textbf{iLog signed} &
    \textbf{iLog data} &
    $\mu$ \textbf{Age} ($\sigma$) &
    \% \textbf{women}\\
    \midrule
    China  & 
    \JLU &
    1,007 &
    69 &
    41  & 
    54  &
    44 &
    19.4 (2.2) & 61\\
    Denmark  & 
    \AAU     & 
    412       & 
    16     & 
    15     & 
    27  & 
    20 &
    28.2 (6.3) & 
    58\\
    India  & 
    \AMRITA  & 
    4,183    &
    141    &
    38  & 
    64  &
    45 &
    21.4 (2.9) & 29 \\
    Italy  &
    \UNITN  &
    5,692   &
    287    &
    215    & 
    263 &
    241 &
    22.1 (3.2) & 56\\
    Mexico  &
    \IPICYT  &
    40   &
    9    & 
    11   & 
    21   & 
    21    & 
    25.2 (4.1) & 33\\
    Mongolia  &
    \NUM     &
    3,972     & 
    214    &
    152    & 
    231 &
    201 & 
    20.0 (3.1) & 65\\
    Paraguay  &
    \UC &
    1,342  & 
    33 & 
    25 & 
    43 &
    28 &
    23.3 (5.1) & 61\\
    UK  & 
    \LSE  &
    1,980  &
    143    & 
    45     &
    88 &
    66 & 
    24.6 (5.0) & 66\\
    \midrule
    \textbf{Total}  &
    & 
    \textbf{18,628}     &
    \textbf{912}   &
    \textbf{542}    &
    \textbf{\nilogusers} &
    \textbf{666} & & \\
    \bottomrule
    \end{tabular}
\end{table}

Study participation is described according to the protocol outlined in \cref{subsec:protocol}.
As shown in \cref{tab:participants}, during the invitation phase, 18,628 students responded to the initial questionnaire, with peak participation at the \UNITN, \AMRITA, and \NUM pilot sites, each registering over 3,000 responses. Most sites saw more than 1,000 students participate, except for \IPICYT and \AAU. The lower participation at these sites is likely due to the recruitment strategies used at \IPICYT and the communication channels adopted at \AAU. In the first case, participants may not have encouraged their acquaintances to take part in the survey, as in the usual snowball sampling procedures. In contrast, in the second case, students might not have been used to checking their emails or visiting the institutional website, so they had not received the communications.

After the invitation phase, up to 350 students from each site were randomly selected for participation in the iLog data collection. This selection process considered factors such as gender and the students' enrolled departments to ensure a representative sample. This approach aimed to address potential biases in the data that might arise from differences in lesson schedules across departments, which could influence students' daily behavior.

Among the invited participants, \nilogusers students installed the iLog app. \cref{tab:participants} details the number of students who signed into the app and those who actively provided data. The number of participants varies across universities, with the highest participation rates observed at \NUM and \UNITN and the lowest at \AAU and \IPICYT. These differences are also evident in the second and third questionnaires, distributed exclusively to previously selected participants. The uneven participation across pilot sites could be due to several factors, such as smartphone compatibility with the iLog app, individual inclinations and preferences, and variations in the research protocols and engagement methods (detailed in \cref{sec:limits}). However, the dataset remains consistent and suitable for high-quality analysis, as shown in \cref{sec:discussion}, presenting a unique opportunity for methodological exploration and analysis. These analyses can examine participant profiles in relation to survey participation, as well as response rates and response times (as outlined below), offering valuable insights for optimizing future data collection.

\begin{figure}[t]
\begin{minipage}[c]{0.47\linewidth}
\includegraphics[width=\textwidth]{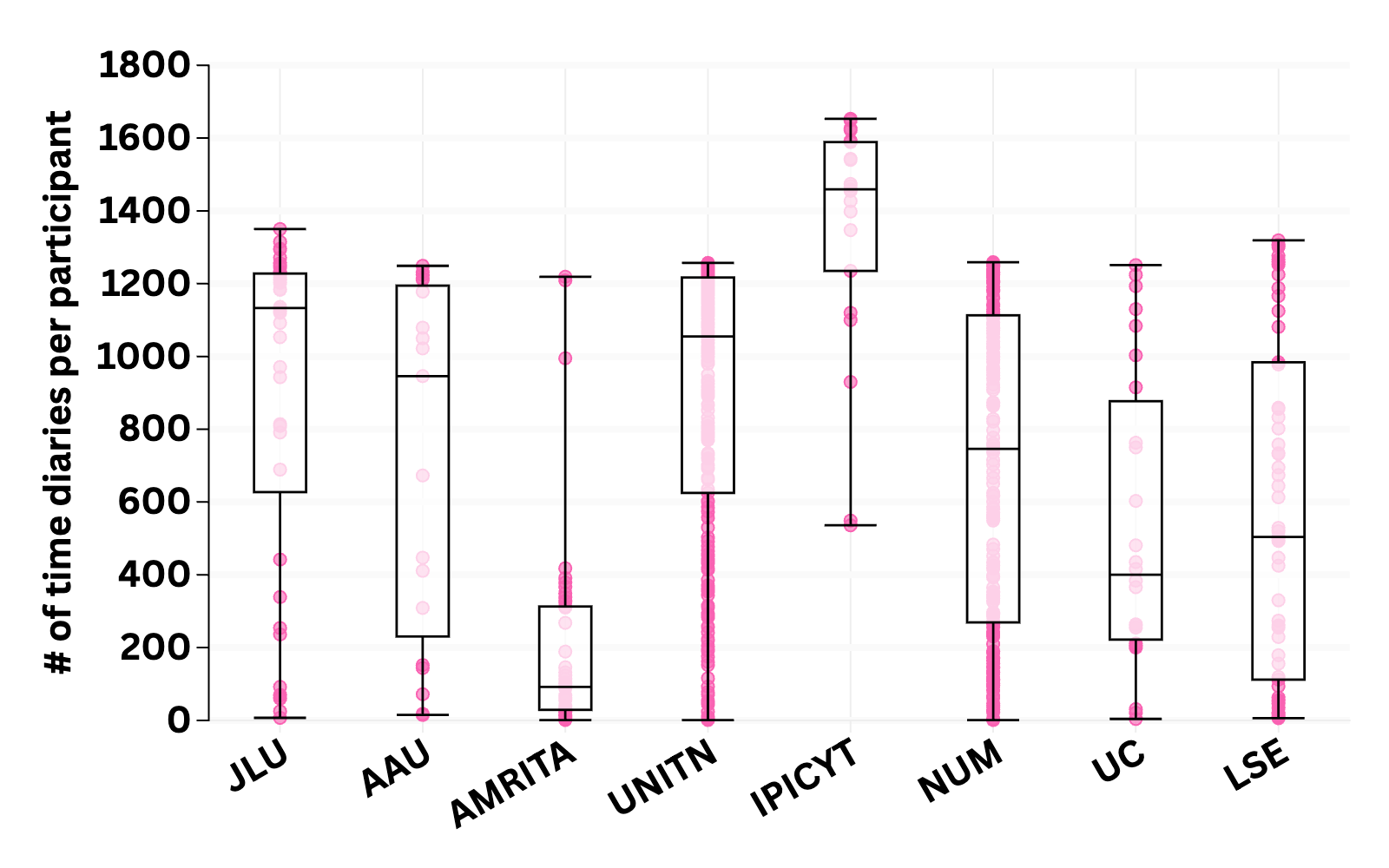}
 \caption{Distribution of participants based on the number of daily diaries completed. Each dot represents a participant.}
 \Description{Box plot showing the participants distribution for each university based on the number of answered time diaries. The mean value is around 1200 in \JLU, 1000 in \AAU, 100 in \AMRITA, 1100 in \UNITN, 1500 in \IPICYT, 800 in \NUM, 400 in \uc and 500 in \LSE.}
 \label{fig:val:td:answers}
\end{minipage}
\hfill
\begin{minipage}[c]{0.47\linewidth}
\includegraphics[width=\linewidth]{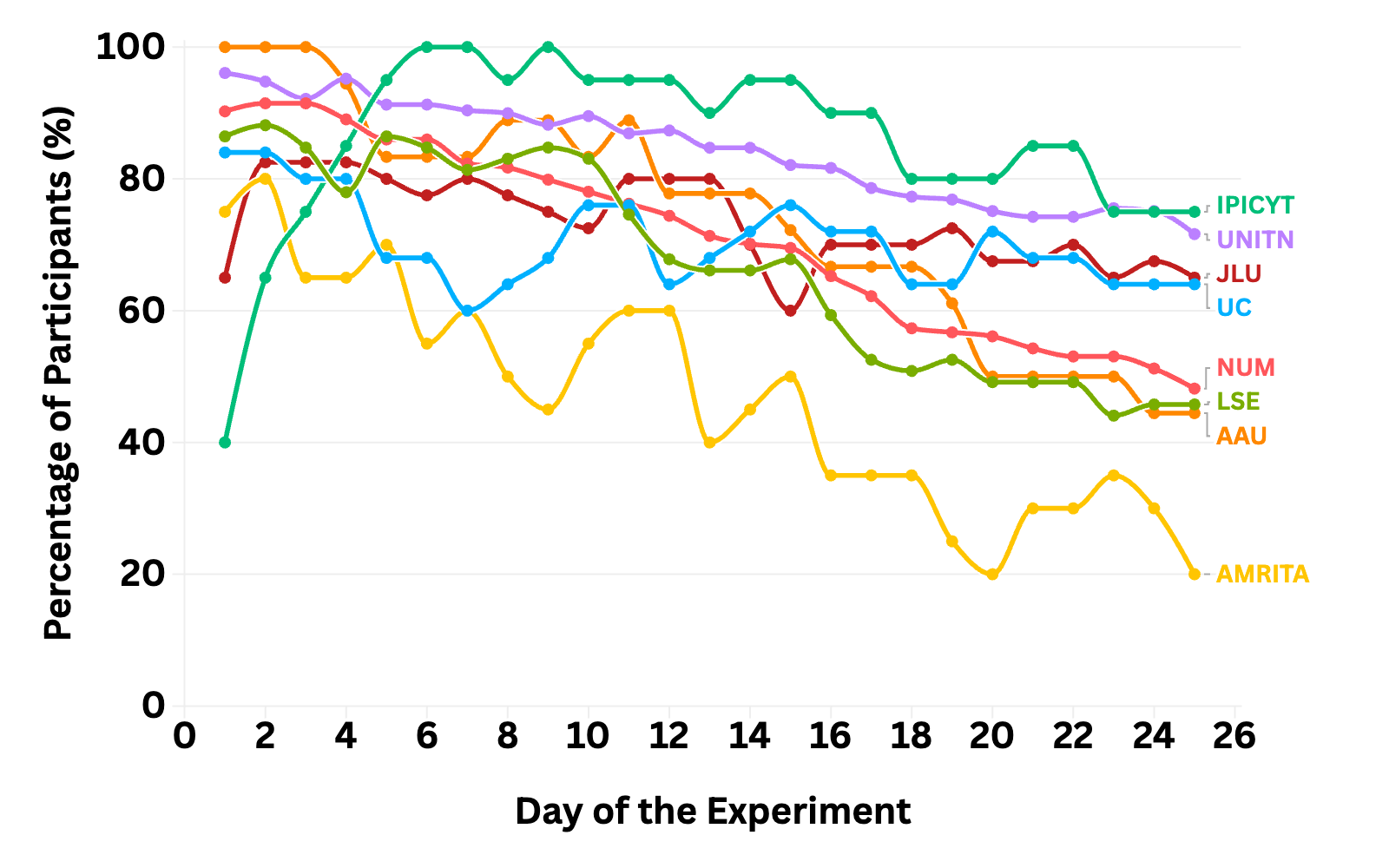}
\Description{Line plot showing for each university the decreasing percentage of users providing data over one month of experiment. The x-axis reports time, and the y-axis the percentage.}
\caption{Percentage of participants at each pilot site providing sensor data for each day.}
\label{fig:val:dropout}
\end{minipage}%
\end{figure}

Having introduced participant engagement at various stages, \cref{fig:val:td:answers} and \cref{fig:val:dropout} illustrate the response rates and sensor data provision during the intensive longitudinal survey, as discussed in the following sections.

\subsection{Response Rate in Time Diaries} \label{subsec:dailyresponses}

\cref{fig:val:td:answers} shows the number of responses to the daily questions outlined in \cref{sec:td}. We expected approximately 1400 responses per participant throughout the study, including morning and evening questions, time diaries, and snack questions. In all pilots (except \AMRITA), 40\% of participants provided an average of 20 responses daily, equating to roughly ten hours of annotated sensor data each day. Notably, the daily response rate is significantly higher at \UNITN and \NUM, where over 150 and 130 participants, respectively, provided more than 20 responses per day. \JLU and \IPICYT also had high participation rates; however, the latter's increased response rate was partly due to the study configuration (see \cref{subsec:ils}), which included more questions than other pilot sites.

Variations in response rates can be attributed to several factors, including adaptations to the survey protocol mentioned above, as well as technical issues and participant characteristics. Some participants did not receive all notifications correctly due to smartphone settings, connectivity issues, or server management. While these technical problems affected only a small fraction of the data, they may have impacted participant engagement.

Furthermore, the data collection process still required the participants to be considerably careful and organized, despite measures such as break options and maintaining active notifications for 12 hours (see \cref{subsec:ils}). Personal factors, such as interest and curiosity (like the appeal of tracking one’s activities to increase self-awareness), as well as tendencies toward procrastination or stress, may have influenced response consistency.

Ultimately, each student’s consistency of participation mainly varied due to dropout effects. While some students ceased involvement after a few days, others maintained consistent engagement throughout the data collection period.

\begin{figure}[hbt]
    \includegraphics[width=0.95\textwidth]{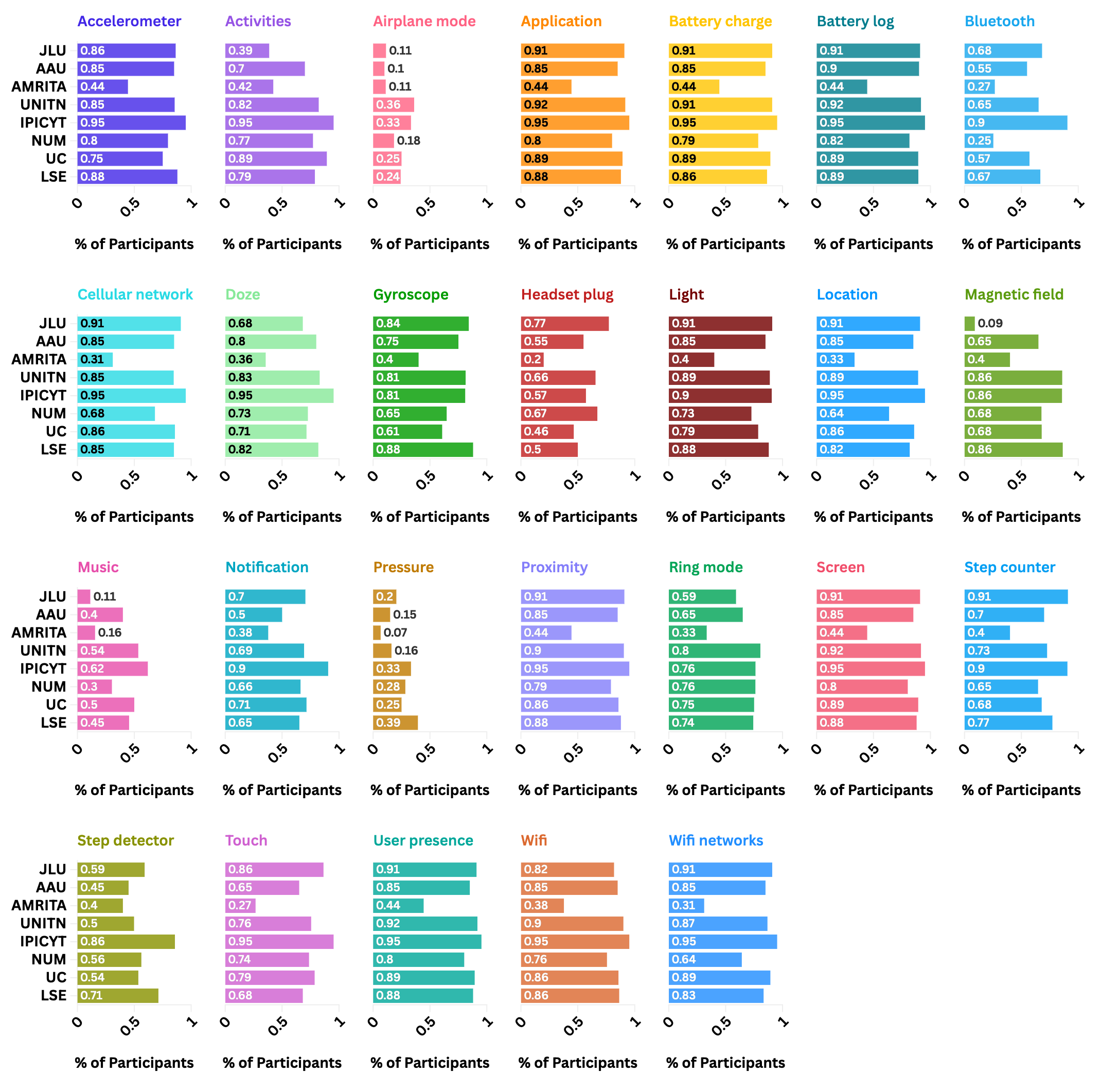}
    \Description{One horizontal barplot for each sensor, each bar representing the percentage of participants with data in one university.}
    \caption{Average percentage of participants that provided sensor data for each site.}
    \label{fig:sensors:participantstats}
\end{figure}

\subsection{Sensor Data Provision}

Sensor data collection adhered to state-of-the-art methods, ensuring that studies based on this dataset can achieve high replicability. We also present details on dropout rates in \cref{fig:val:dropout}, along with the percentage of participants who provided data compared to the total participants with at least one recorded event, shown in \cref{fig:sensors:participantstats}\footnote{A comprehensive description of the sensors and their collection frequency is available in \cref{tab:sensor-list,tab:sensor}.}.

\cref{fig:val:dropout} illustrates the percentage of participants providing sensor data over time with respect to the total number of participants who contributed data. The most notable decrease occurs at \AMRITA, where participation drops from 80\% on the first day of data collection to less than 25\% by the final day. On average, the dropout rate across sites is approximately 35\%. The initial increase observed at \JLU and \IPICYT results from participants who joined the data collection after the start date.

For each sensor, \cref{fig:sensors:participantstats} shows the percentage of participants who contributed in relation to all participants with at least one recorded event (see also the “iLog data” column in \cref{tab:participants}). A lower percentage does not necessarily indicate low data quality. For instance, the number of participants reporting data from on-change sensors, such as airplane mode or music playback, is influenced by the actual usage of these smartphone features. Sensors collected at fixed intervals, like the accelerometer and application usage, tend to have a higher number of participants with data. In contrast, sensors like pressure capture data from a smaller subset of participants.

\noindent
In addition to dropouts, variations in sensor data provision may be attributed to application or device issues and the lack of support for specific hardware sensors on some devices. Participants could also disable any sensor directly from the app at any time for privacy considerations. For instance, a participant could disable GPS tracking if visiting a location they preferred to keep private (e.g., a place of worship or a hospital). Additionally, some sensors may have been disabled to conserve battery life. Together, these factors contribute to the observed instances of missing data. Hence, although this privacy feature resulted in some missing data, we believe this design significantly contributed to attracting a higher number of participants who felt comfortable providing data over an entire month.

\subsection{Diversity in Everyday Life}

\begin{figure}
    \centering
    \begin{subfigure}[b]{0.33\textwidth}
    \includegraphics[width=\textwidth]{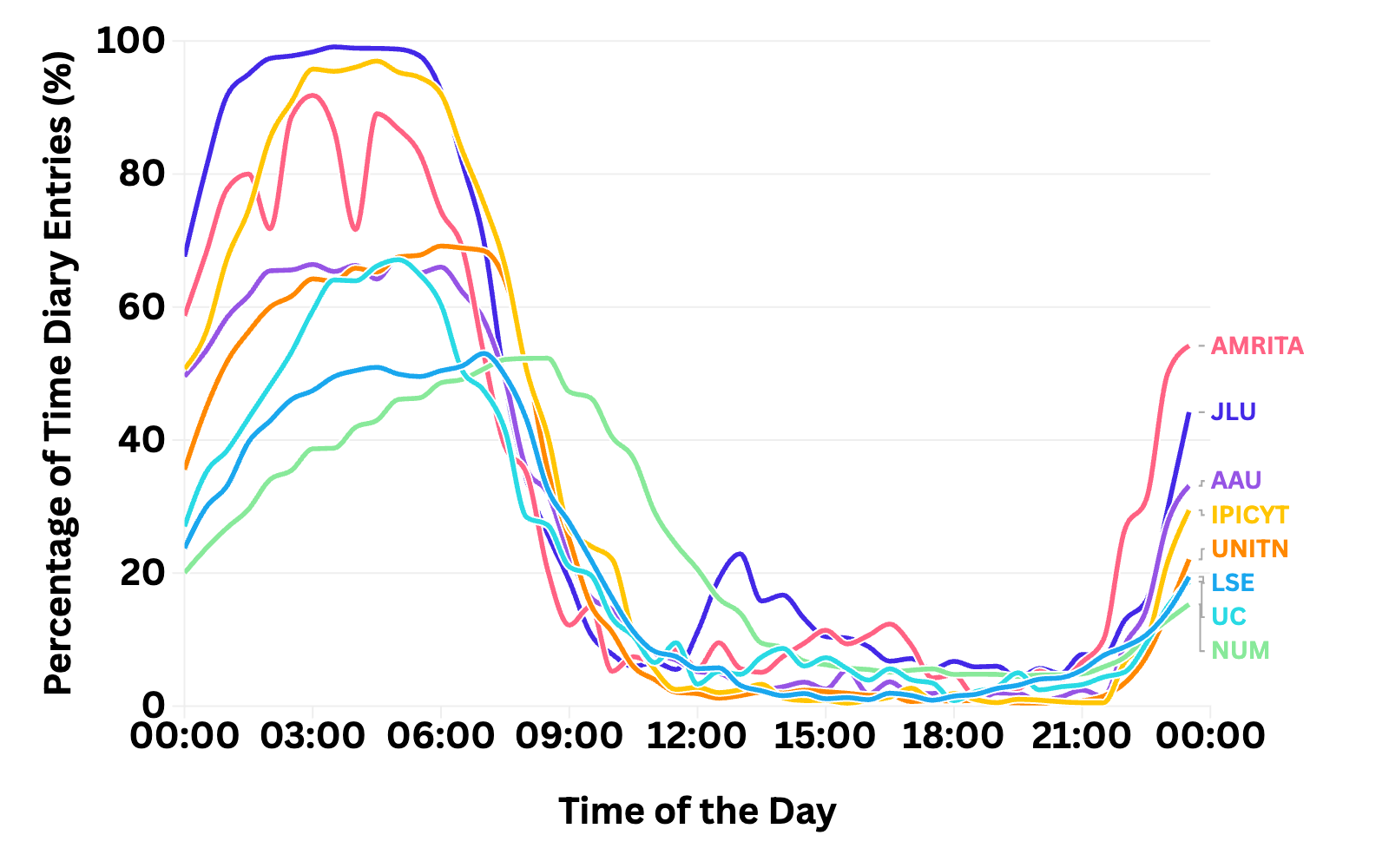}
    \caption{Sleeping}
    \label{fig:val:td:disthours:sleeping}
    \end{subfigure}
    \begin{subfigure}[b]{0.33\textwidth}
    \includegraphics[width=\textwidth]{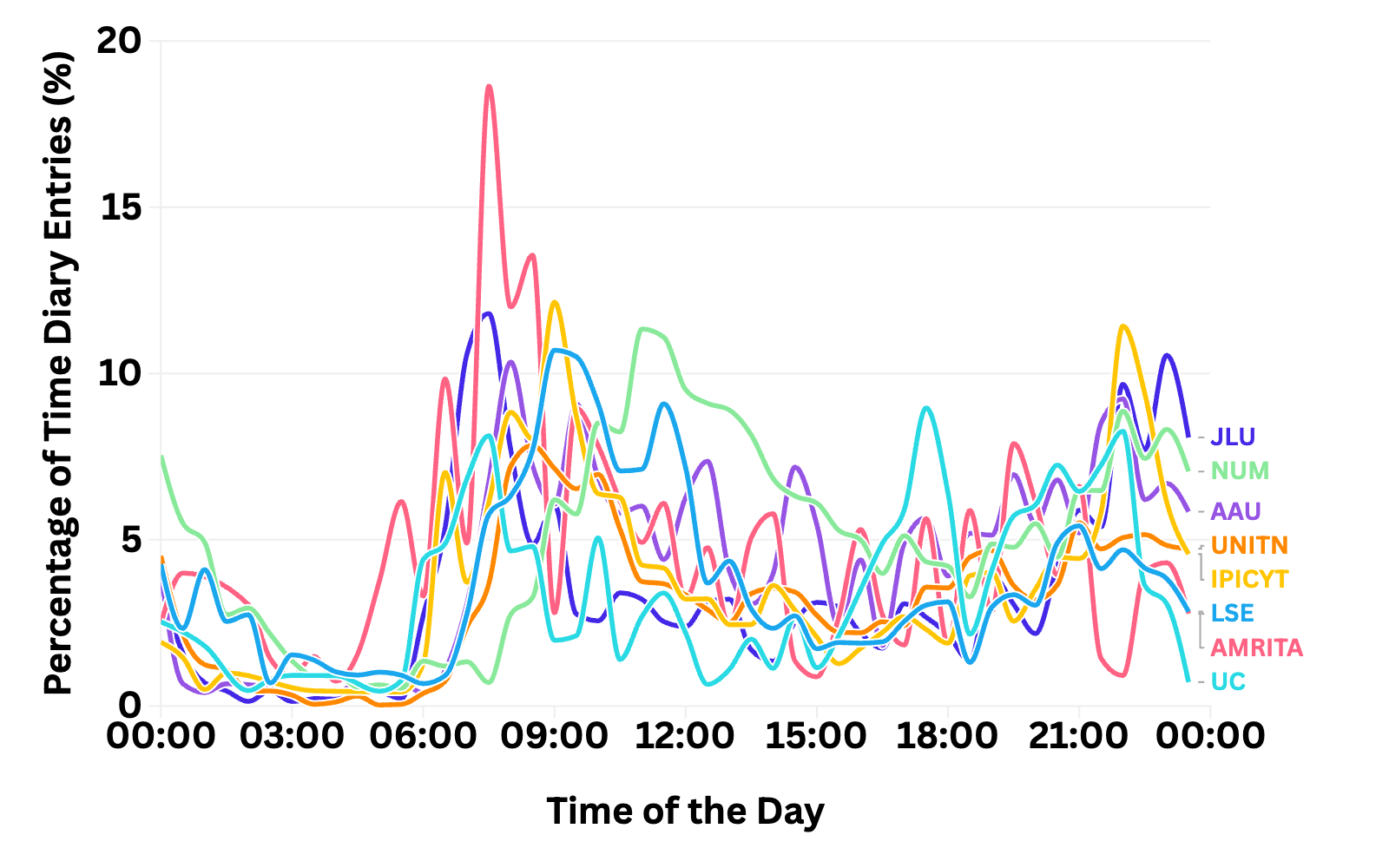}
    \caption{Eating}
    \label{fig:val:td:disthours:eating}
    \end{subfigure}
    \begin{subfigure}[b]{0.33\textwidth}
    \includegraphics[width=\textwidth]{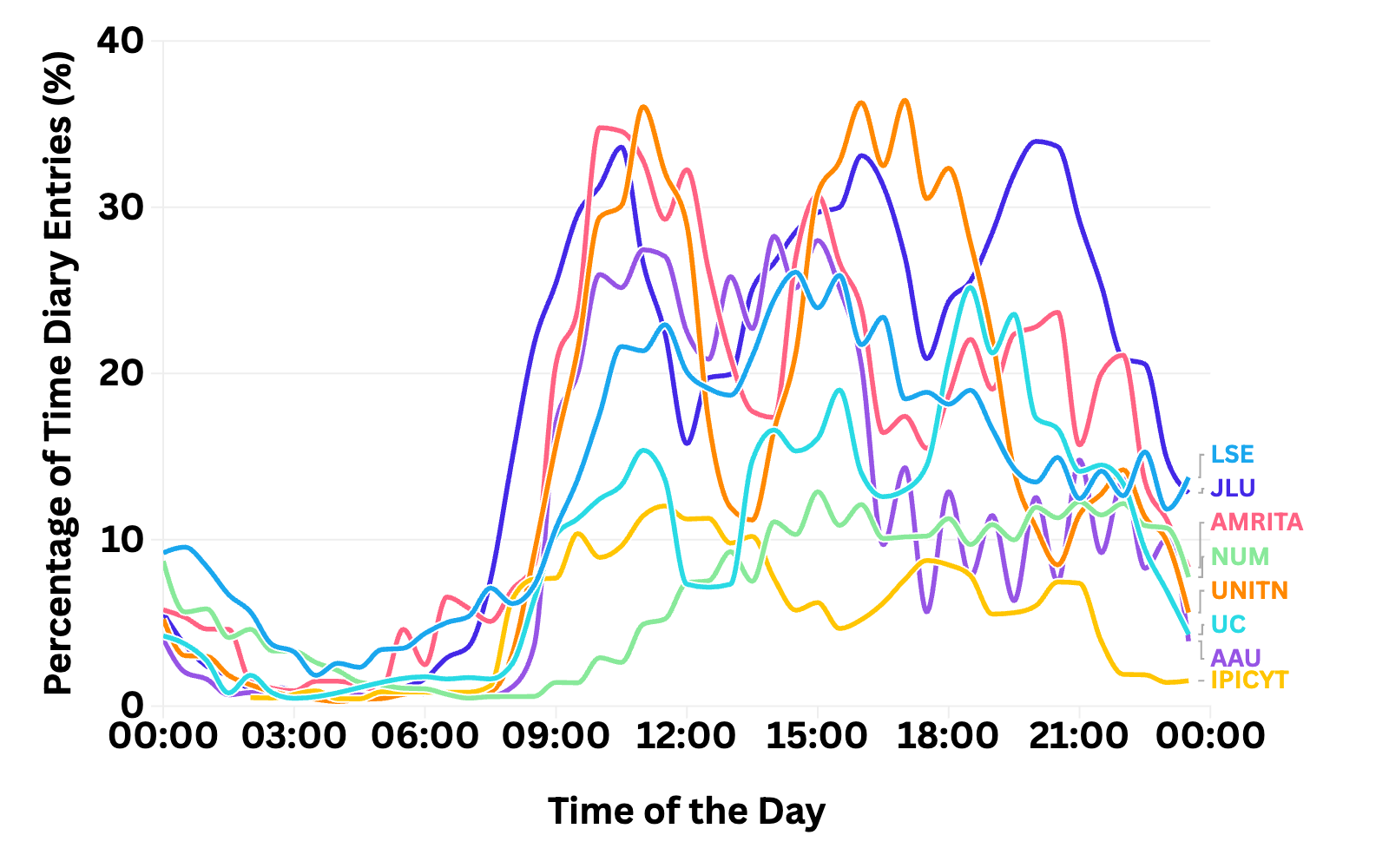}
    \caption{Studying}
    \label{fig:val:td:disthours:studying}
    \end{subfigure}
    \Description{Line plot for each university showing the percentage of participants reporting the sleeping, eating and studying activity over the day. The x-axis is time and the y-axis the percentage.}
    \caption{Proportion of time diary reports during different hours of the day.}
    \label{fig:val:td:disthours}
\end{figure}

\begin{figure}
    \centering
    \begin{subfigure}{0.49\textwidth}
        \includegraphics[width=\textwidth]{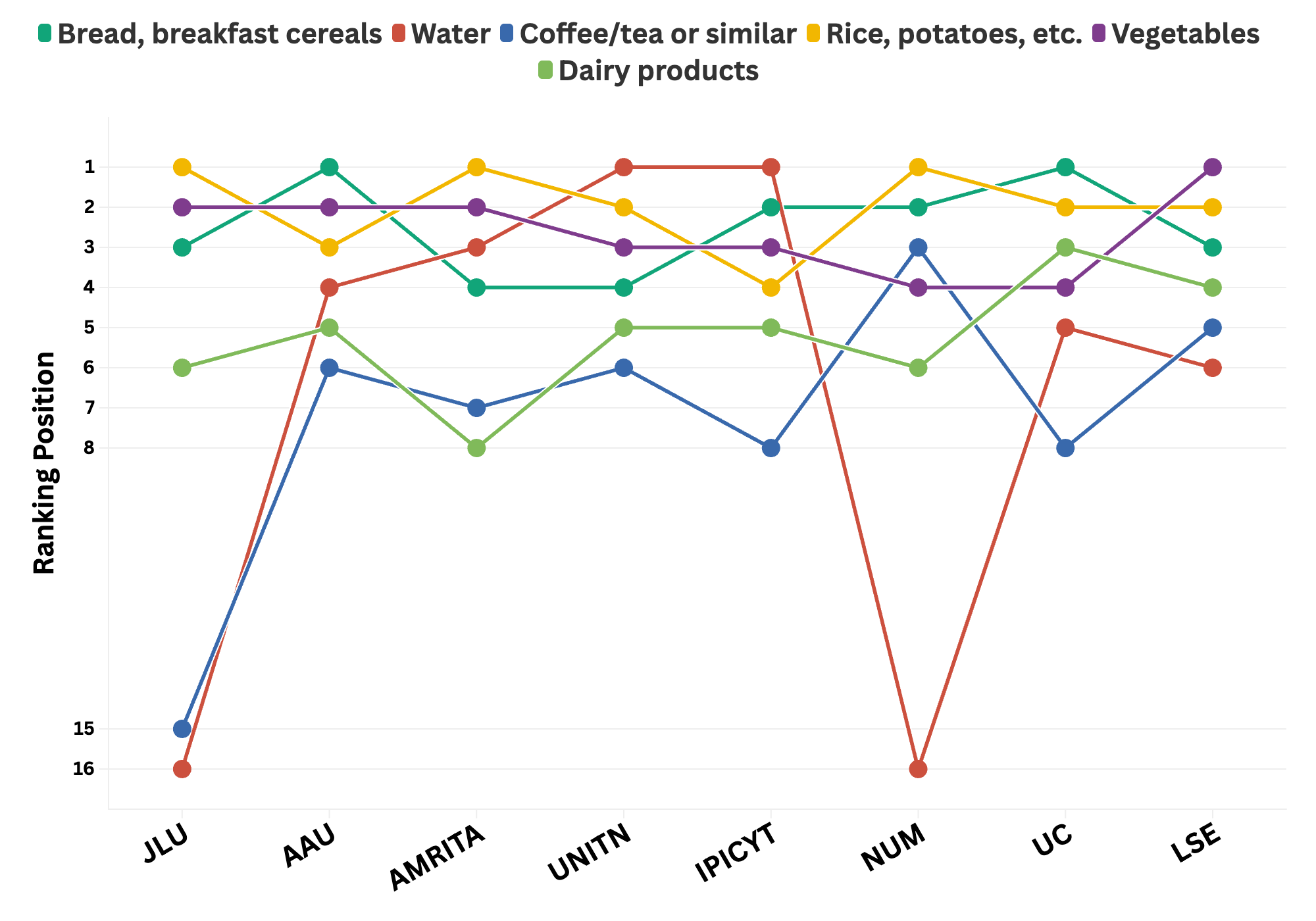}
        \caption{Foods}
        \label{fig:val:td:food:food}
    \end{subfigure}
    \begin{subfigure}{0.49\textwidth}
        \includegraphics[width=\textwidth]{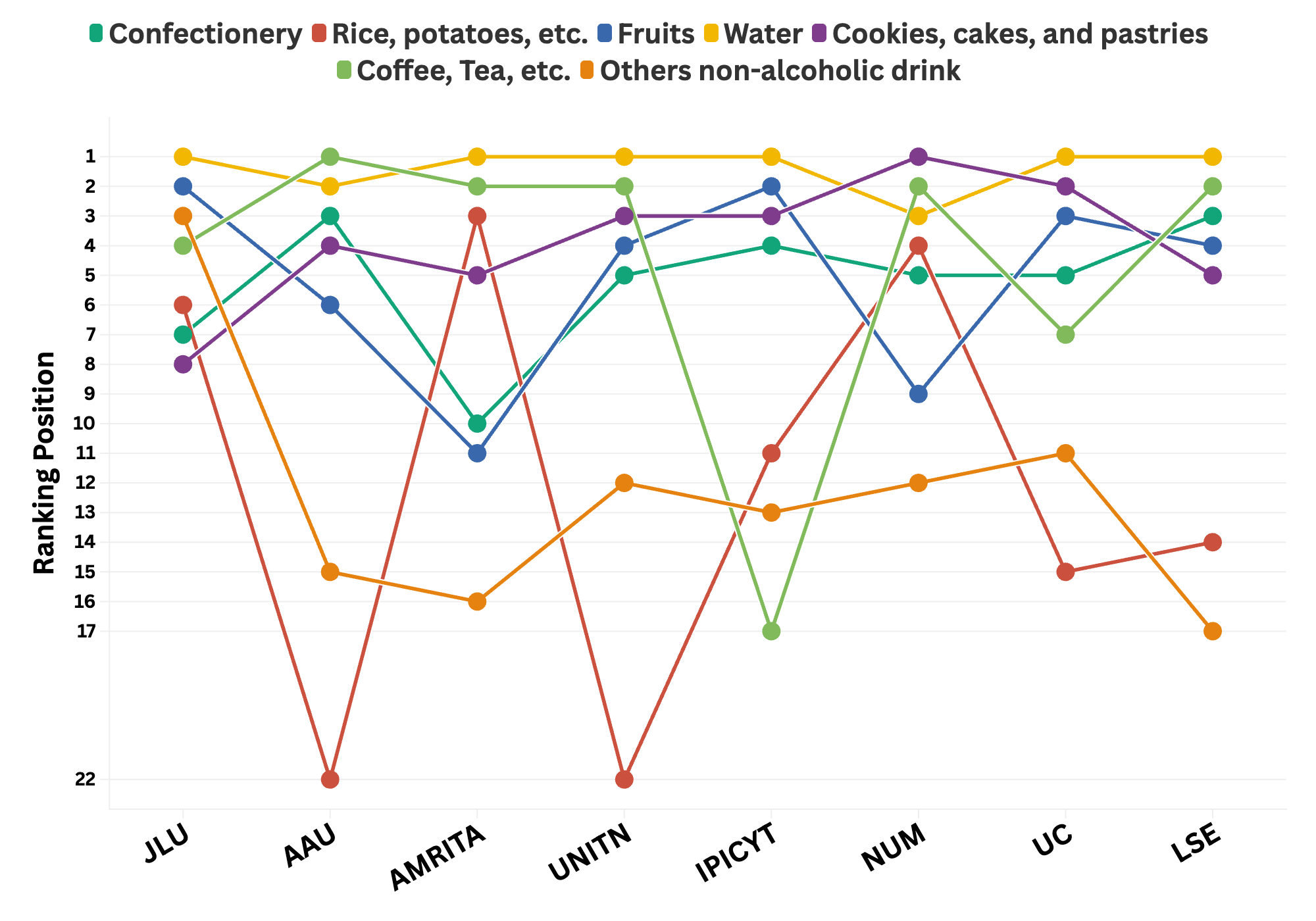}
        \caption{Snacks}
        \label{fig:val:td:food:snack}
    \end{subfigure}
    \Description{The two plots show the ranking position of each food and snack in each university. X-axis represents universities, y-axis the ranking position.}
    \caption{Ranking comparison of the three most consumed foods and snacks.}
    \label{fig:val:td:food}
\end{figure}


To highlight behavioral variations, we analyze the distribution of primary daily activities—sleeping, eating, and studying—across different pilot sites. For instance, \cref{fig:val:td:disthours} displays these differences. Notably, \cref{fig:val:td:disthours:sleeping} shows distinct sleeping patterns between midnight and 7:00 AM, with many students from \JLU resting in the afternoon, despite similar waking times to other pilot sites. Comparing these patterns with study hours (\cref{fig:val:td:disthours:studying}) reveals that study-related annotations account for over 30\% of activities at 8:00 PM in \JLU, compared to just 7\% in \IPICYT. It is also worth noting that sleep-related annotations were primarily captured using an app feature that allowed participants to define sleep time windows for several hours before returning to regular self-reporting in the morning.

The timing of eating activities also varies among universities. As shown in Figure \ref{fig:val:td:disthours:eating}, \AMRITA reports eating spikes around 8:00 AM and 10:00 PM, while \NUM peaks around 11:00 AM and \UC around 6:00 PM. Differences in eating behaviors are further highlighted in Figure \ref{fig:val:td:food}, which ranks the top three foods and drinks reported at each site. For example, water is the most consumed drink at \UNITN, while it ranks only 16th at \NUM and \JLU (refer to Figure \ref{fig:val:td:food:food}). Interestingly, water consistently appears among the top-reported drinks during snack breaks across all sites, as illustrated in Figure \ref{fig:val:td:food:snack}.

\begin{figure}
    \centering
    \begin{subfigure}{0.31\textwidth}
        \includegraphics[width=\linewidth]{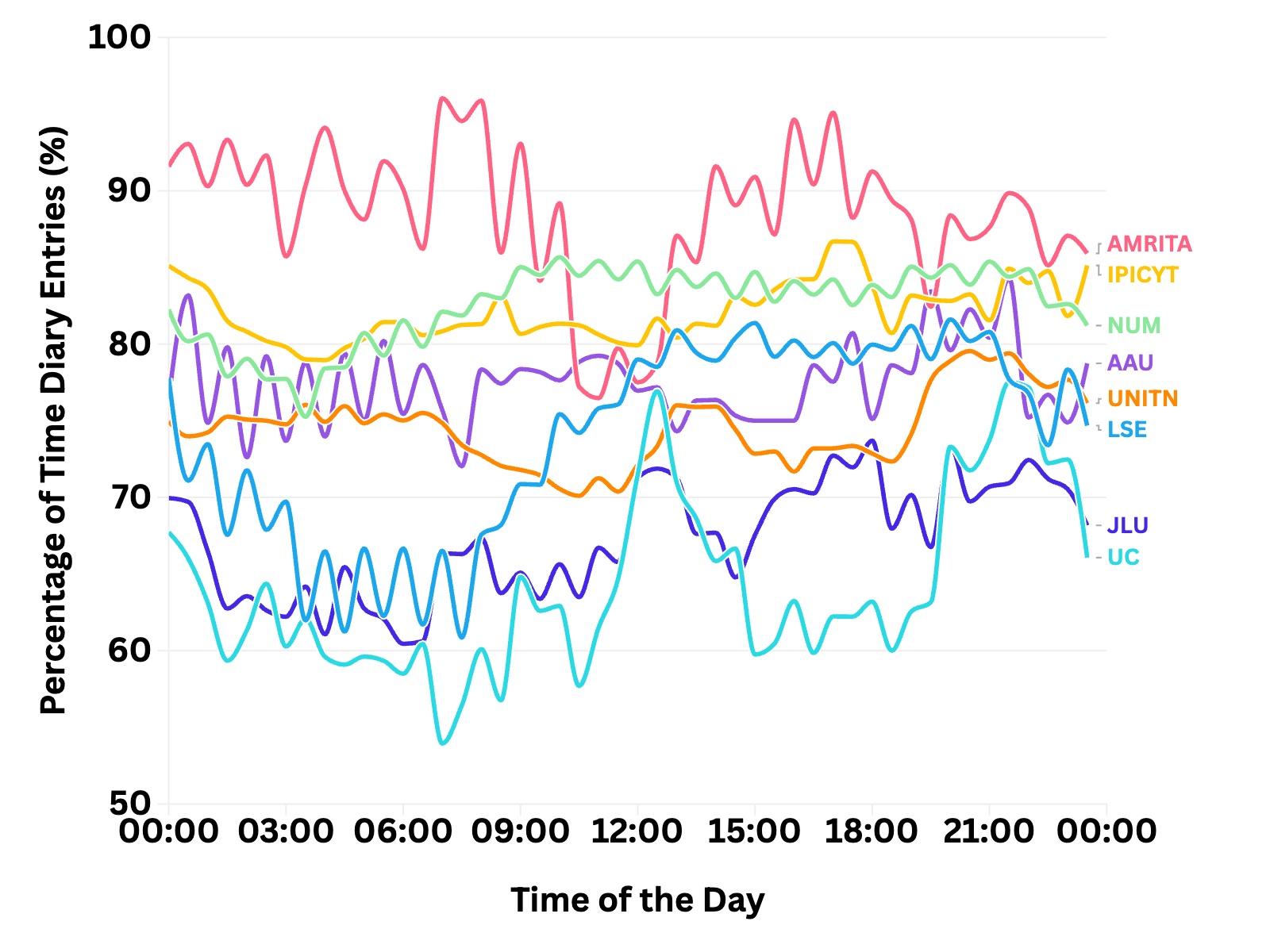}
        \Description{Line plot for each university. The X-axis is the percentage of participants reporting positive mood, and the y-axis is the hours of the day.}
        \caption{Percentage of positive mood reports over the day for each site.}
        \label{fig:mood_location:hourdist}
    \end{subfigure}
    \hfill
    \begin{subfigure}{0.31\textwidth}
        \includegraphics[width=\linewidth]{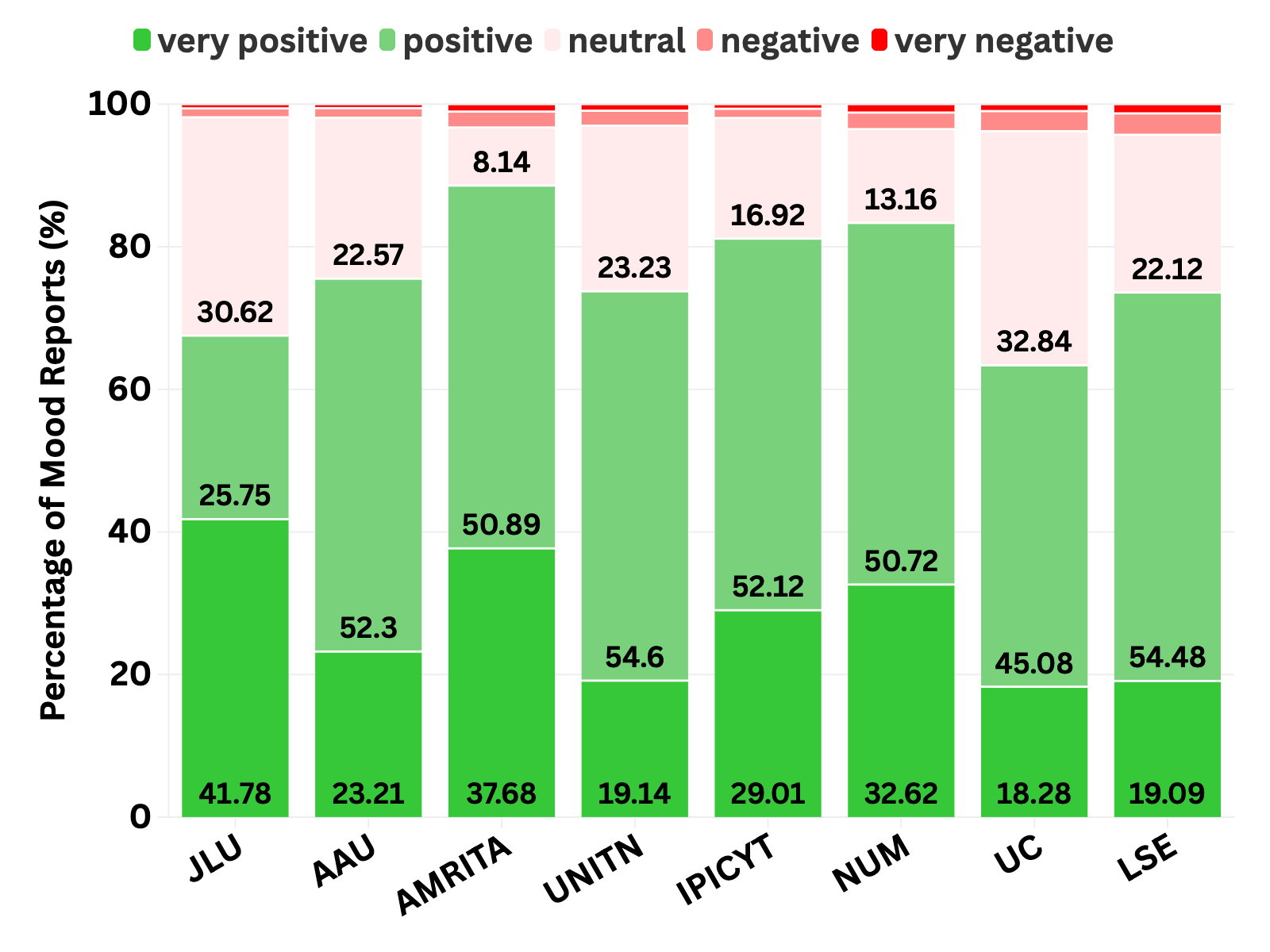}
        \Description{Stacked bar plot reporting the overall mood distribution when the participant is at home, for each university.}
        \caption{Mood distribution when the participants are at home.}
        \label{fig:mood_location:home}
    \end{subfigure}
    \hfill
    \begin{subfigure}{0.31\textwidth}
        \includegraphics[width=\linewidth]{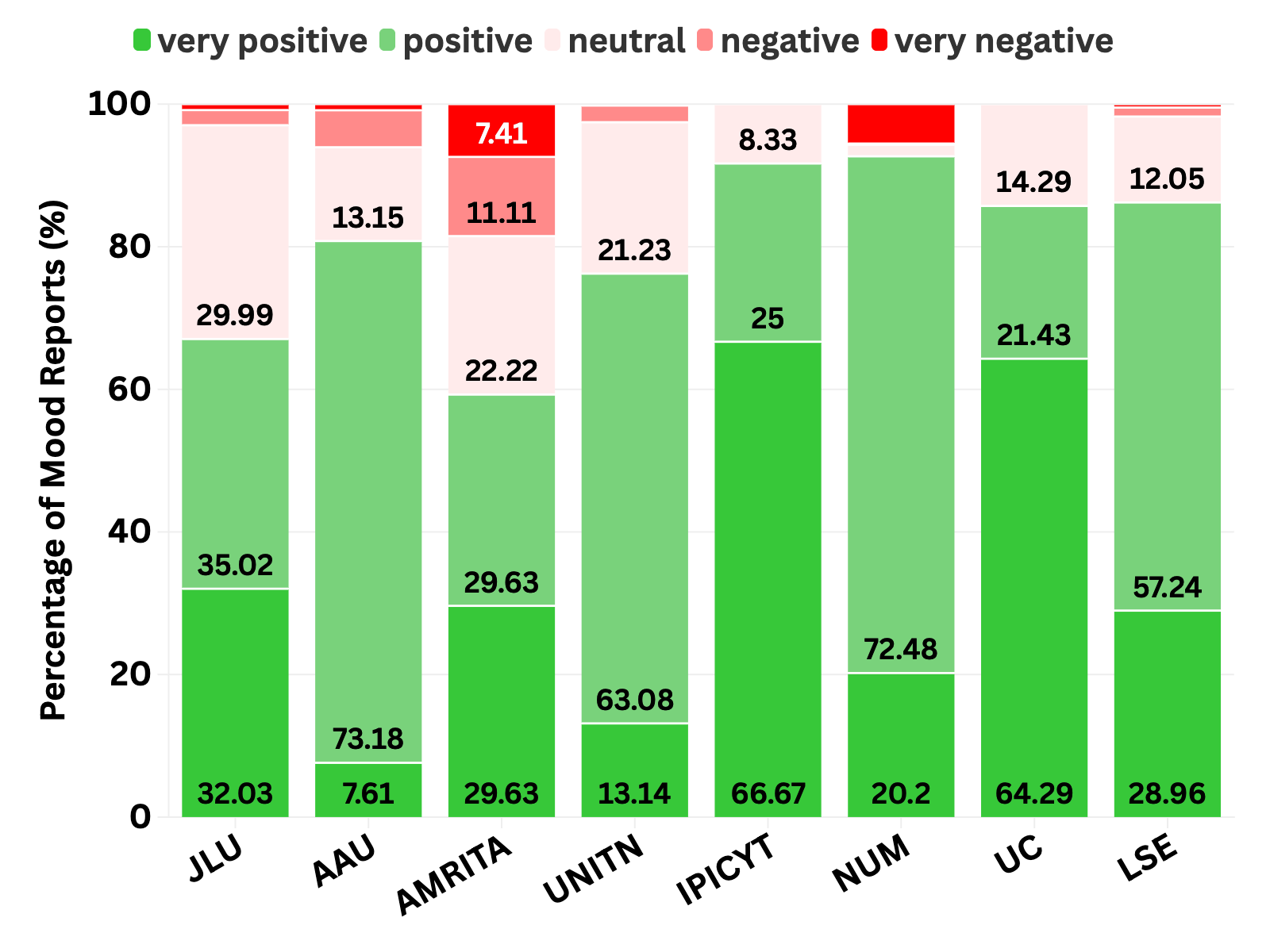}
        \Description{Stacked bar plot reporting the overall mood distribution when the participant is at the university, for each university.}
        \caption{Mood distribution when the participants are at the university.}
        \label{fig:mood_location:uni}
    \end{subfigure}
    \caption{Mood ratings distribution.}
    \label{fig:mood_location}
\end{figure}

Mood reporting adds an intriguing layer to this study, allowing researchers to investigate how mood ratings vary in similar contexts across different universities (what was captured is, actually, a facet of mood called Valence). This has been discussed in detail by Meegahapola et al. \cite{meegahapola2023generalization}, who used this dataset for one of their studies). Figure \ref{fig:mood_location:hourdist} illustrates changes in positive mood over the day. While trends appear similar across locations, absolute percentages vary, with \UC reporting the lowest mood levels and \AMRITA the highest. Participants’ personal rating criteria also influence these differences. Furthermore, mood differences are noticeable depending on location: distinct distributions of mood ratings are evident when participants are at home (e.g., apartments, gardens, or relatives' homes) compared to university spaces (e.g., classrooms and libraries), as shown in \cref{fig:mood_location:home} and \cref{fig:mood_location:uni}.

Research questions regarding phone usage and time spent on activities such as internet browsing, chatting, or gaming can be examined by analyzing application usage. \cref{tab:application} ranks applications based on the number of participants using them. As expected, social networking and messaging apps are the most common across universities, with \JLU students, for instance, favoring local applications such as WeChat.

This section offers only an initial glimpse into the potential analyses and research questions that can be pursued by combining and reshaping the data. A more comprehensive discussion can be found in \cref{sec:discussion}.

\begin{table}
\centering
\small
\caption{Ranking of each institution's five most used mobile applications.}\label{tab:application}
\begin{tabular}{llllll}
\toprule
 & \textbf{1st} & \textbf{2nd} & \textbf{3rd} & \textbf{4th} & \textbf{5th} \\
\midrule
\JLU & WeChat & QQ - Tencent & Taobao & Bilibili & Google Messages \\
\AAU & Youtube & Facebook Messenger & Spotify & Google Chrome & Google Maps \\
\AMRITA & Android settings & Whatsapp & Google Chrome & Android settings & Youtube \\
\UNITN & Whatsapp & Google Gmail & Youtube & Google Chrome & Android settings \\
\IPICYT & Facebook & Whatsapp & Google quick search & Google Chrome & Android settings \\
\NUM & Facebook Messenger & Facebook & Youtube & Android settings & Google Play Store \\
\UC & Google Chrome & Whatsapp & Instagram & Android settings & Youtube \\
\LSE & Whatsapp & Google Chrome & Android settings & Google Play Store & Google Gmail \\
\bottomrule
\end{tabular}
\end{table}

\section{Dataset and Code Availability} \label{sec:availability}
Due to the variety of sources and the sensitive nature of the content, this dataset is made available in a secure environment that complies with GDPR regulations. The main entry point documentation can be accessed in the dataset catalog:

\vspace{1em}
\tikzstyle{background rectangle}=[thin,draw=black]
\begin{center}
\begin{tikzpicture}[show background rectangle]
\node[align=justify, text width=30em, inner sep=1em]{
\url{https://livepeople.disi.unitn.it}
};
\node[xshift=3ex, yshift=-0.7ex, overlay, fill=white, draw=white, above 
right] at (current bounding box.north west) {
\textit{Link to the dataset catalog}
};
\end{tikzpicture}
\end{center}

\noindent
The catalog simplifies the process of finding and selecting datasets of interest by providing metadata and materials to explore their features. \change{The filter options allow to restrict the search on \dataset data.} It also ensures privacy and copyright compliance, fully respecting the ownership of the authors who contributed to its collection. \change{In addition to the catalog, the webpage available at \url{https://datascientia.disi.unitn.it/projects/diversityone/} collects further information.} The descriptions of downloadable datasets, associated documentation, and procedures for dataset requests are detailed below.

\subsection{Downloadable Datasets and Bundles}

\begin{table}[tb]
    \centering
    \small
    \caption{\change{The size of each bundle of data in Gigabytes, with uncompressed sizes in brackets, and bundle names as they appear in the catalog are presented. Synchronic and diachronic interaction refers to questionnaires and time diaries, respectively.}}\label{tab:sizes}

\begin{tabular}{lllllllll}
\toprule
\textbf{Bundle name} & \textbf{\JLU} & \textbf{\AAU} & \textbf{\AMRITA} & \textbf{\UNITN} & \textbf{\IPICYT} & \textbf{\NUM} & \textbf{\UC} & \textbf{\LSE} \\
\midrule
App-usage & $<$ 0.1 (0.3) & $<$ 0.1 (0.2) & $<$ 0.1 (0.2) & 0.3 (2.8) & $<$ 0.1 (0.4) & 0.1 (2.3) & $<$ 0.1 (0.3) & $<$ 0.1 (0.7) \\
Connectivity & $<$ 0.1 (2.5) & 0.1 (5.7) & $<$ 0.1 (0.4) & 0.6 (26.6) & $<$ 0.1 (1.2) & $<$ 0.1 (3.7) & $<$ 0.1 (2.8) & 0.4 (16.8) \\
Device-usage & $<$ 0.1 (0.4) & $<$ 0.1 (0.3) & $<$ 0.1 (0.2) & 0.5 (4.4) & $<$ 0.1 (0.2) & 0.2 (2.3) & $<$ 0.1 (0.4) & $<$ 0.1 (0.9) \\
Environment & 0.6 (5.6) & 0.3 (2.6) & 0.1 (1.4) & 5.2 (44.5) & 0.6 (6.4) & 3.3 (35.2) & 0.5 (5.4) & 1.5 (12.1) \\
Motion & 1.2 (12.4) & 1.9 (18.0) & 1.3 (13.2) & 18.0 (174.8) & 4.0 (36.8) & 17.1 (161.3) & 1.8 (19.9) & 7.0 (61.3) \\
Position & 0.1 (1.7) & 0.6 (5.2) & 0.6 (5.5) & 12.7 (104.8) & 1.3 (10.2) & 6.8 (54.7) & 0.8 (7.6) & 2.9 (23.0) \\
Synchronic int. & $<$ 0.1 & $<$ 0.1 & $<$ 0.1 & $<$ 0.1 & $<$ 0.1 & $<$ 0.1 & $<$ 0.1 & $<$ 0.1 \\
Diachronic int. & $<$ 0.1 & $<$ 0.1 & $<$ 0.1 & $<$ 0.1 & $<$ 0.1 & $<$ 0.1 & $<$ 0.1 & $<$ 0.1 \\
\midrule
Total & 2.1 & 2.9 & 2.1 & 37.3 & 6.0 & 27.7 & 3.3 & 12.0 \\
\bottomrule
\end{tabular}

\end{table}

The resources are organized and made available separately considering the full dataset size (approximately 94GB in Parquet format) and GDPR’s minimization principle—which mandates that data must be adequate, limited, and relevant for analysis. Researchers can request access to basic datasets from a specific pilot site (e.g., time diaries collected at \UNITN) or a combination of datasets from multiple pilot sites. To streamline dataset selection, we have created thematic bundles that group data commonly used together for main research purposes. For example, the motion bundle includes all motion sensor data relevant to activity recognition studies, while another bundle, combining questionnaires, time diaries, and location data, is tailored for \cancel{social science research}\change{studying social interactions}. \change{The catalog lists both datasets containing one single sensor and bundles.} \cref{tab:sizes} reports the available bundles (\cref{subsec:ils} details the sensors they contain), and Appendix~\ref{app2:sensors} reports the complete list of sensors in each bundle. \change{Any combination of bundles and single sensors can be downloaded.} All datasets are provided in Parquet format\footnote{Apache Parquet \url{https://parquet.apache.org/}}, an efficient storage format with high compression.
Access to the entire \dataset dataset is also available upon request. 
No source code needs to be made available for the release of this dataset. Future benchmarked datasets derived from our raw dataset will be made available, respecting privacy and copyright, along with code for pre-processing and machine learning-based modeling.

\subsection{Metadata and Documentation}
Each single sensor dataset and bundle is accompanied by metadata and comprehensive documentation that outline content, size, format, and other relevant information.
The catalog user can search the datasets through the metadata values such as the acronym, data collection location, and type of bundle or dataset. The metadata for each dataset and bundle includes:
%
    \textit{(i)} A technical report detailing the data collection process;
    \textit{(ii)} Dataset metadata and a codebook with summary statistics;
    \textit{(iii)} Data collection information and links to related projects, including articles published using the dataset;
    \textit{(iv)} Documentation and procedures for dataset requests.

The catalog enhances the findability, accessibility, and reusability of the dataset in several ways. First, it provides an efficient search method for specific data within a secure institutional environment. Second, it includes concise descriptions that guide users through the available resources, along with detailed information on each dataset, including variable values and labels. Lastly, the technical report and dataset descriptions enable users to analyze the dataset and replicate the data collection process for their own research needs.

\subsection{Dataset Request Process}

\change{We outline the procedure to request the data.
First, the user retrieves the identifiers of the bundle or single sensor data of interest from the metadata shown on the catalog for each sensor and bundle.
Second, the metadata also includes a link to the request form, designed together with legal and privacy experts to comply with GDPR and privacy regulations. Interested users affiliated with a research institution can request bundles and datasets by listing the identifiers selected in the previous step and presenting a research proposal. Upon completing the form, users submit their request to the catalog manager using the email address specified in the metadata.} Detailed eligibility criteria and required information are available on the catalog website.
Third, once approved, users must sign a \change{Terms} and License Agreement. Key licensing terms include: \textit{(i)} datasets \cancel{may be}\change{are} used exclusively for research purposes; \textit{(ii)} redistribution of the datasets is prohibited; \textit{(iii)} datasets cannot be publicly shared (e.g., on a website); and \change{\textit{(iv)} any attempt to reverse engineer any portion of the data or to re-identify the participants is strictly forbidden and could constitute unlawful processing of personal data. Finally, the catalog manager sends the instructions for downloading the dataset.}

\section{Discussion} \label{sec:discussion}
\begin{table}[t]
    \centering
    \caption{\change{Summary Recommendations for Future Cross-Country Smartphone-Based Data Collection.}}
    \label{tab:recommendations}
    \begin{tabular}{p{0.02\linewidth} p{0.85\linewidth}}
        \toprule

        \multicolumn{2}{l}{\textbf{\textsc{Recommendation \#1: Core Protocol Design}}} \\
        & 
        Develop a standardized core protocol that allows flexible local adaptations. Consult with local experts early to validate instruments, and pilot-test before
        large-scale deployment. \\

        \arrayrulecolor{Gray}
        \midrule

        \multicolumn{2}{l}{\textbf{\textsc{Recommendation \#2: Cultural Sensitivity}}} \\
        & 
        Involve local stakeholders to ensure cultural norms are respected.
        Conduct small-scale pilots to refine sensitive questions, and leverage cultural experts for precise translations.  \\

        \arrayrulecolor{Gray}
        \midrule

        \multicolumn{2}{l}{\textbf{\textsc{Recommendation \#3: Privacy and Ethics}}} \\
        & 
        Create a modular privacy framework compliant with international (e.g., GDPR) and local regulations. Use adaptable templates and regularly consult legal experts to monitor changes.\\

        \arrayrulecolor{Gray}
        \midrule

        \multicolumn{2}{l}{\textbf{\textsc{Recommendation \#4: Data Anonymization}}} \\
        & 
        Adopt multi-level anonymization (e.g., personal identifiers, location data) to balance privacy with data utility. Offer multiple data versions (e.g., “RoundDown” GPS vs. “POI” GPS) to accommodate different re-identification risks.\\

        \arrayrulecolor{Gray}
        \midrule

        \multicolumn{2}{l}{\textbf{\textsc{Recommendation \#5: Incentive Strategies}}} \\
        & 
        Tailor incentives to local economic conditions and cultural preferences. Collaborate with local partners to choose monetary vs. non-monetary rewards for maximum engagement.\\

        \arrayrulecolor{Gray}
        \midrule

        \multicolumn{2}{l}{\textbf{\textsc{Recommendation \#6: Technological Adaptation}}} \\
        & 
        Design or select survey tools that can operate both online and offline. Plan for restricted services (e.g., Google) and ensure solutions work in diverse mobile ecosystems.\\

        \arrayrulecolor{Gray}
        \midrule

        \multicolumn{2}{l}{\textbf{\textsc{Recommendation \#7: Participant Engagement}}} \\
        & 
        Maintain transparent communication through notifications, reminders, and helpdesks. Offer support channels that address technical or procedural issues quickly without biasing responses.\\
        \arrayrulecolor{black}
        \bottomrule
    \end{tabular}
\end{table}

\subsection{Data Collection -- Lessons Learned and Recommendations for the Future}\label{subsec:lessons_design}

\change{In Section 3, we detailed our core design decisions for gathering cross-country \dataset dataset. Building on those design considerations, this section focuses on the broader lessons learned during implementation, compares our experiences with other local data collection practices, and discusses how to adapt these insights for future studies. We consolidate all recommendations into Table~\ref{tab:recommendations}.}

\change{Collecting smartphone data across diverse cultural and regulatory contexts offered an invaluable opportunity to observe how different countries and research teams navigate ethical, logistical, and cultural considerations. Our collaboration revealed several common threads shared with local researchers as follows. First, \textit{Cultural Adaptation:} while our study required adjustments in survey content, incentive structures, and communication styles, local researchers often faced similar challenges. For instance, local studies had to modify language and question framing to ensure sensitivity to cultural or religious norms. Our experiences echo these practices but highlight the importance of designing a standardized core protocol that can be flexibly adapted without sacrificing data comparability. Second, \textit{Privacy and Regulatory Compliance:} many local research groups reported ongoing adaptations to meet international guidelines (e.g., GDPR) while conforming to national regulations. Our approach—having a core GDPR-compliant framework and then tailoring it to meet local legal expectations—aligned with similar multi-country studies in the literature. This comparison highlights the benefits of centralized documentation that can be adjusted to local requirements. Third, \textit{Technological Infrastructure and Offline Solutions:} the need to accommodate varying degrees of internet connectivity and restricted services was not unique to our study; many local researchers rely on hybrid models (online/offline) for data gathering. Our development of an “offline” version of the \texttt{iLog} app is congruent with other best practices in low-connectivity settings, such as storing data locally and syncing at scheduled intervals.}


\change{Hence, Table~\ref{tab:recommendations} brings together our main recommendations and the current reflections. Each item links directly to a stage of the study—from protocol design to data anonymization—and highlights how researchers can adapt these suggestions to their own contexts. By integrating our field experiences, we observed that successful cross-country data collection depends on a core set of adaptable protocols. We hope these consolidated recommendations will guide future multi-country studies in designing, implementing, and scaling smartphone-based data collection responsibly and effectively.}

\subsection{Dataset analysis -- Lessons Learned and Recommendations for the Future}\label{subsec:lessons_analysis}

The analysis of \dataset  across various studies has revealed critical insights into the influence of local context, country-specific nuances, and multi-modal data streams on the effectiveness of mobile sensing models.
Presented below are key lessons learned from these studies, each accompanied by a recommendation for future research. 

\begin{table}[t]
\centering
\small
\caption{\change{A summary of previous studies that utilized the \dataset dataset is presented in the table. The acronyms used are as follows: Target/s Inferred refers to the primary aspect that was focused in the particular study; CS indicates that country-specific models were trained; MC denotes that multi-country models were trained by combining data from multiple countries; CA refers to the examination of a country-agnostic setting, focusing on the generalization performance of models across countries; and PER indicated that model personalization was examined; SL refers to supervised learning; FT/RT refers to transfer learning with fine-tuning or re-training for classic machine learning models to achieve personalization; and UDA refers to unsupervised domain adaptation.}}\label{tab:prev_studies_summary}
\begin{tabular}{llllllll}
\toprule
 & \textbf{Venue and Year} & \textbf{Target/s Inferred}                                         & \textbf{CS}  & \textbf{MC}   & \textbf{CA} & \textbf{PER} & \textbf{Methods} \\
\midrule
Bouton–Bessac et al. \cite{bouton2022your} & PervasiveHealth '22 & Complex Daily Activities              & $x$ & $\checkmark$  & $x$  & $\checkmark$ & SL, FT/RT \\
Meegahapola et al. \cite{meegahapola2023generalization} & IMWUT '23 & Mood              & $\checkmark$ & $\checkmark$  & $\checkmark$  & $\checkmark$ & SL, FT/RT \\
Assi \& Meegahapola et al. \cite{assi2023complex} & CHI '23 & Complex Daily Activities  & $\checkmark$ & $\checkmark$  & $\checkmark$  & $\checkmark$  & SL, FT/RT \\
Kammoun et al. \cite{kammoun2023understanding} & ICMI '23 & Social Context when Eating  & $\checkmark$ & $\checkmark$  & $x$      & $\checkmark$  & SL, FT/RT \\
Girardini et al. \cite{girardini2023adaptation} & EPJ DataScience '23 & Daily routines and Covid       & $\checkmark$ & $x$  & $x$  & $\checkmark$  & SL, FT/RT \\
Meegahapola et al. \cite{meegahapola2024m3bat} & IMWUT '24 & Mood, Social Context       & $\checkmark$ & $x$      & $\checkmark$  & $x$   & SL, UDA   \\
Mader \& Meegahapola et al. \cite{mader2024learning} & CHI '24 & Social Context         & $\checkmark$ & $\checkmark$  & $\checkmark$  & $\checkmark$  & SL, FT/RT \\

\bottomrule
\end{tabular}
\end{table}

\subsubsection{Country-specific Models over Multi-country Models}

Studies consistently demonstrate that models trained on country-specific data outperform those trained on multi-country datasets. For tasks such as mood inference \cite{meegahapola2023generalization} and complex daily activity recognition \cite{assi2023complex}, country-specific models achieved higher accuracy, with AUROC scores ranging from 0.76 to 0.98 when using partially personalized, hybrid models. Conversely, generic models not adapted to specific countries often struggled when applied to new regions, failing to capture unique data distributions that reflect localized contexts. This gap emphasizes the critical importance of country-specific patterns and nuances in daily routines and sensor data, which vary widely and directly influence data interpretation. These findings are in line with the findings of Khawaja et al. \cite{khwaja2019modeling}. 

Moreover, these insights shed light on limitations faced by previous multi-country sensing studies, such as those conducted by Servia-Rodriguez et al. \cite{servia2017mobile}. By pooling data from multiple countries, these models averaged out region-specific patterns, reducing predictive power and capacity to capture unique, localized behaviors. For example, in the study by Meegahapola et al. \cite{meegahapola2023generalization}, location and movement data were highly predictive of mood in certain countries but held minimal relevance in others, underscoring regional differences in daily routines and smartphone usage. This diversity indicates that a one-size-fits-all approach overlooks essential cultural and contextual factors crucial for accurate behavior inference, reinforcing the need for adaptable, localized models.

Computer vision models provide a valuable contrast here. Convolutional neural networks (CNNs) and vision transformers excel at capturing features in images—textures, edges, and spatial hierarchies—regardless of geographic origin. These models generalize effectively across multi-country visual datasets by learning local and global image features, excelling when regional content differences are less impactful on performance. However, smartphone time-series data captures complex, region-specific behavioral nuances that current time-series models struggle to interpret consistently across geographies. Developing models that can capture local nuances in behavioral data while generalizing across regions remains crucial for advancing global smartphone sensing applications.

\subsubsection{Value of Hybrid Model Personalization}

\change{Typically, models developed for various prediction tasks are tested on a different population to evaluate their performance, reflecting real-world scenarios. These are referred to as population-level models. 
However, prior work \cite{zhang2024reproducible, meegahapola2023generalization, assi2023complex} in ubiquitous computing has shown that such models often underperform, especially for highly subjective or complex inferences. Therefore, the population-level model should either be fine-tuned or re-trained for the target individual using a small portion of that individual's data (known as hybrid models), or, if sufficient data are available and it is computationally feasible, a model can be trained solely on the target individual's data (fully personalized models).
} Research using \dataset dataset has shown that even modest personalization using hybrid models enhances performance in behavioral inference tasks such as mood detection \cite{meegahapola2023generalization}, complex activity recognition \cite{assi2023complex}, and social context inference \cite{mader2024learning}. This aligns with findings from other datasets, such as those studied by Zhang et al. \cite{zhang2024reproducible}, which further validate the value of hybrid models. Hybrid models, which combine general population data with user-specific features, improve accuracy by balancing broad patterns with individual nuances. Fully personalized models can yield the highest accuracy but require extensive user-specific data, making them resource-intensive and less scalable. 
Hybrid models offer a practical compromise, delivering strong performance with fewer data demands per user, better accommodating individual behavioral variations.

For example, in mood detection, hybrid models incorporating individual app usage and movement data can capture unique variations, improving accuracy compared to purely generic models. This hybrid approach demonstrates that while general models detect broader trends, they often miss critical individual subtleties that partially personalized models can capture, providing a practical strategy for improving model relevance and precision.

\subsubsection{Advancing Techniques in Multimodal Integration for Mobile Sensing}

Integrating diverse data streams—such as activity, location, Bluetooth, app usage, device use, and WiFi— enhances behavioral inference by capturing a broader spectrum of contextual and activity patterns. Existing research using \dataset has largely relied on foundational multimodal techniques, often without leveraging advanced integration methods tailored to the high-dimensional, diverse data sources typical in mobile sensing. A recent study applying an unsupervised domain adaptation model, M3BAT \cite{meegahapola2024m3bat}, explored multimodal settings, demonstrating that domain adaptation can help mitigate distributional shifts across modalities, achieving up to a 12\% improvement in AUC on specific classification tasks. However, much remains to be done to develop models capable of handling the complexities of multimodal data in mobile sensing, particularly when accounting for regional, cultural, and individual behaviors.

Smartphone sensor data is shaped by regional routines, cultural practices, and individual habits, requiring models to adapt dynamically to context-dependent patterns across data streams. Temporal transformers and adversarial training approaches tailored to multimodal data offer promising directions, though these are not yet widely applied in mobile sensing. Future research could benefit from multimodal architectures that use independent branches for each sensor modality, allowing each data type to be processed in context while adapting to shifting behavioral patterns. The next generation of multimodal models in mobile sensing should aim to improve upon initial efforts by adopting advanced architectures that facilitate robust, context-aware fusion of diverse data sources. Such models would enable richer, more generalized applications, supporting domains like health monitoring, behavior analysis, and activity recognition across diverse populations.

\subsubsection{The Challenge of Distribution Shift}

In cross-country smartphone sensing, distribution shifts in data impact the performance of models trained in one region when applied in another. Recent studies highlight that sensor data distributions vary widely between countries, introducing substantial challenges for model generalization. For instance, Meegahapola et al. \cite{meegahapola2024m3bat} observed high Cohen’s-d values, up to 1.0, across various sensing modalities such as location and step count. These large effect sizes reveal pronounced shifts in data distributions that hinder a model's ability to transfer effectively across geographic contexts. For example, GPS patterns that indicate social activities in one country may reflect commuting in another, necessitating region-specific adaptations to accurately interpret similar data.

While substantial progress has been made in addressing distribution shifts in computer vision and natural language processing \cite{wang2018deep, ge2023domain}, domain adaptation in multimodal mobile sensing is still an emerging field \cite{chang2020systematic, meegahapola2024m3bat, wu2023udama}. As mentioned earlier, the M3BAT framework by Meegahapola et al. employed a multi-branch adversarial training approach to tackle distribution shifts in multimodal sensor data, achieving performance gains of up to 12\% in AUC for classification tasks and a 0.13 MAE reduction in regression tasks. By creating separate branches for features with varying distribution shifts, M3BAT tailored adaptation strategies to each modality, enhancing model robustness. Moreover, Wu et al. \cite{wu2023udama} also applied domain adaptation for multimodal data, however, without considering the multimodality. Xu et al. \cite{xu2023globem} used a domain generalization algorithm, again without considering the multimodality. However, all the above methods are still nascent, addressing only a subset of multimodal data shifts. 

\begin{table}[t]
    \centering
    \caption{Summary Recommendations for Future Research Based on the Dataset Analysis.}
    \label{tab:analysis_recommendations}
    \begin{tabular}{p{0.02\linewidth} p{0.85\linewidth}}
        \toprule

        \multicolumn{2}{l}{\textbf{\textsc{Recommendation \#8: Country-Specific vs.\ Multi-Country Models}}} \\
        
        & 
        Country-specific or regional models could be better for initial deployment in diverse regions, as compared to multi-country models. Tailoring models to capture unique behavioral patterns and cultural nuances in countries enhances accuracy without full personalization. \\

        \arrayrulecolor{Gray}
        \midrule
        
        \multicolumn{2}{l}{\textbf{\textsc{Recommendation \#9: Hybrid Model Personalization}}} \\
        
        & 
        Adopt a hybrid strategy for model personalization by combining general population data with user-specific inputs to capture both local and individual behavioral nuances. This approach achieves a balance between broad generalization and user-specific relevance, enhancing model performance while managing data demands effectively. \\

        \arrayrulecolor{Gray}
        \midrule
        
        \multicolumn{2}{l}{\textbf{\textsc{Recommendation \#10: Advanced Multimodal Integration}}} \\
        
        & 
        Develop advanced multimodal integration techniques for behavioral models in mobile sensing. Multi-branch architectures may provide an effective way forward by facilitating context-aware fusion of diverse data sources. \\

        \arrayrulecolor{Gray}
        \midrule
        
        \multicolumn{2}{l}{\textbf{\textsc{Recommendation \#11: Addressing Distribution Shifts}}} \\
        & 
        Implement domain adaptation techniques that account for cross-country
distribution shifts in mobile sensing. Consider using multi-branch adversarial training to tailor model responses to region-specific patterns, improving generalization across diverse data distributions. \\

        \arrayrulecolor{Gray}
        \midrule

        \multicolumn{2}{l}{\textbf{\textsc{Recommendation \#12: Label Quality}}} \\
        & 
        Prioritize objective labels for tasks where label quality critically impacts model performance. For subjective data, explore advanced preprocessing techniques such as outlier regularization, attention mechanisms, and weighted averaging to reduce noise, thereby improving label consistency and model accuracy. \\

        \arrayrulecolor{black}
        \bottomrule
    \end{tabular}
\end{table}

\subsubsection{Effect of Label Quality on Model Performance}

The quality of labels has been shown to profoundly influence model performance after transer learning, especially in mobile sensing applications where labels can vary widely in terms of subjectivity \cite{zeni2019fixing,bontempelli2020learning,wu2023udama}. A study using \dataset \cite{meegahapola2024m3bat} revealed that models trained on subjective labels, such as self-reported mood, tend to exhibit lower consistency and reliability compared to models trained on more objective labels, such as social context. Although both labels are silver standard \footnote{\change{In machine learning, the term “silver-standard” refers to labels or annotations that, due to uncertainty, bias, or other sources of noise, offer lower accuracy or reliability than “gold-standard” labels. Despite these limitations, silver-standard labels remain valuable and are commonly employed in inference tasks. They may be derived from a variety of less rigorous methods, including automated algorithms, self-reports, surrogate measures, or indirect observations. \cite{wu2023udama, meegahapola2024m3bat, dy2023domain}}} because they are self-reported, social context is less subjective and mood is more subjective, leading to various behaviors for models. This difference suggests that, where possible, researchers should aim to use objective and gold standard labels \footnote{\change{In machine learning, “gold-standard” refers to the most accurate and reliable ground truth labels or annotations. They are often derived from highly trusted methods—such as expert manual annotations, precise measurements, or comprehensive and well-established criteria. However, producing these gold-standard labels can be challenging, time-consuming, and costly, as it typically requires specialized expertise, rigorous data collection, and meticulous validation \cite{wu2023udama, meegahapola2024m3bat, dy2023domain}.}} to enhance model reliability and accuracy. However, for tasks that inherently rely on subjective input—like mood, depression, stress or perceived energy levels---it is crucial to develop preprocessing methods that can transform subjective data into more standardized, reliable labels. Techniques such as regularizing outliers, employing weighted averages for conflicting reports, or integrating machine learning methods like attention mechanisms to focus on higher confidence data points can help mitigate the impact of noise in subjective labels.

By integrating these lessons with advanced modeling techniques, researchers can enhance the robustness and scalability of mobile sensing applications in diverse cultural and geographic contexts. These recommendations aim to guide future work toward more accurate, context-aware models, ultimately supporting broader real-world adoption in areas such as health monitoring, behavior prediction, and social context inference.

\subsection{Limitations}\label{sec:limits}

Although we observed a high response rate for questionnaires at many pilot sites, participation in the intensive longitudinal survey varied between countries. For example, \UNITN and \NUM had exceptionally high participation numbers, while \AMRITA faced a notably high dropout rate, with some cases reporting completion rates below 40\%. Despite achieving strong results in certain areas of data collection during the challenges of the COVID-19 pandemic, we have identified several critical reflections on the study's progress to better contextualize the disparities observed in the collected data.

First, although the methodology included a local adaptation phase, communication challenges arose during the design and preparation stages. These issues likely stemmed from the involvement of experts outside the social sciences and/or ubiquitous computing, which may have disadvantaged those less familiar with questionnaire design or intensive longitudinal surveys \cite{helm2023diversity}. To foster a more inclusive and diversity-aware approach in future data collections involving interdisciplinary teams, it would be beneficial not only to share and adapt data collection materials but also to provide methodological guidance that informs design strategies across fields, as mentioned in \cref{subsec:lessons_design}. 

Second, it is evident that the recruitment strategy yielded varying results across pilot sites. The decision to use a uniform approach—sending an email invitation to all students at each institute—was based on methodological considerations regarding sampling strategies, data standardization, and comparability. However, cultural differences and distinct socialization processes led to divergent response rates. For example, while some sites recorded over 5,000 responses to the invitation questionnaire, others received very few. This suggests that future communication campaigns should better account for cultural diversity by utilizing a range of channels and enriching promotional materials, possibly incorporating videos, interviews, or participant testimonials to engage broader audiences.

Third, although sensor data collection remained consistent across countries, participants outside of \AMRITA submitted a high volume of sensor data meeting state-of-the-art standards (see \cite{assi2023complex}). Nevertheless, high dropout rates persisted throughout data collection, a common issue in intensive longitudinal surveys. Various factors contribute to this problem, many of which relate to incentives (discussed further below) and respondent burden. For instance, delays in providing support (especially if app-related issues arose) and communication challenges likely impacted participants’ continued engagement. Although break options were designed to alleviate respondent burden (outlined in Section \ref{subsec:ils}), these were not always utilized, perhaps due to expectations for plug-and-play apps or an overwhelming amount of materials, including privacy documentation. To address this, a single repository or website where participants could easily access all relevant information, supported by tutorials and videos, would likely improve usability and engagement, which is to be explored in future work. 

Fourth, disparities in the proposed incentives appear to have influenced data collection success in some contexts. This finding aligns with Singer’s \cite{singer2013use} assertion that fixed monetary compensation generally outperforms gifts or prize draws in motivating participation unless no incentive is provided. Therefore, it is advisable to offer fixed monetary incentives or a set of equivalent alternatives, such as university credits combined with vouchers for students. Additionally, exploring incentives such as digital badges, personalized messages, or gamification elements may be worth considering to sustain long-term engagement.

Finally, the COVID-19 pandemic presented an unforeseen circumstance that invalidated some initial design choices, complicating efforts to ensure ecological validity. In response, we adjusted the survey as needed and conducted a follow-up study at the end of 2021 to examine the pandemic's impact on student’s lives through the lens of smartphone sensors, providing the possibility of comparative insights for post-pandemic contexts in future studies, even though the later dataset is not released as part of \dataset. This second will be released soon, hopefully within 2024. \cite{girardini2023adaptation} studies the effects of Covid-19 by comparing a subset of DiversityOne, limited to UniTN, with a dataset collected earlier on, in 1918, on the same population.

\section{Conclusion}\label{sec:conclusion}
Human behavior varies not only by time of day but also across individuals, countries, and cultures, introducing layers of complexity that most existing smartphone sensing and self-report datasets fail to capture. Current datasets are often limited to specific countries in the Global North, small participant samples, or a narrow set of sensors, which restricts their ability to reflect cross-country behavioral variations and nuanced sensor data patterns. To address this gap, we introduce \dataset, a dataset designed to combine passive smartphone sensor data with self-reported labels at multiple granularities. \dataset is built on a robust interdisciplinary methodology that incorporates standards from computer science, behavioral sciences, social practice theory, and intensive longitudinal surveys via time diaries. This approach supports machine learning models in inferring various everyday aspects, including mood, activities, social context, and eating habits. Data were collected across eight countries from \nilogusers college students over four weeks, encompassing \nsensors smartphone sensors and more than 350,000 self-reports, positioning \dataset as one of the largest and most diverse datasets of its kind, that is publicly available. Additionally, \dataset provides raw sensor data, offering researchers flexibility in processing and usage for diverse analytical purposes. As a significant contribution of this paper, \dataset is publicly released with a comprehensive description of the study design and data collection process, highlighting key insights, lessons learned, and future research directions toward diversity- and privacy-aware studies. We hope this dataset will serve as a valuable resource for the community, enabling studies across multiple disciplines, including ubiquitous computing, human-computer interaction, and machine learning.


\section{Author contributions}

The names order is by the contribution of the institution and, inside each institution, by the contribution of the individuals. As such, the order of names does not necessarily reflect the importance of the contribution of single individuals. The roles of the authors are presented by their initials as follows:
\begin{itemize}
    \item \textit{Study management}: F.G., I.B., A.D.G., M.B.; 
    \item \textit{Study design}: F.G., I.B., G.G., A.D.G., M.B., R.C.A., L.M., D.G.P.; 
    \item \textit{Technical support}: M.R., M.B., R.C.A., L.J.M.;
    \item \textit{Data Collection}: M.B., R.C.A., M.R., A.D.G, P.K., A.G., A.C., G.G., S.S., M.B., L.C., A.H., H.X., D.S., S.D., C.N., S.R.C., A.R.M.;
    \item \textit{Data Preparation and correction}: A.B., R.A.A., I.K., R.C.A., I.B., M.B., D.G.P., L.M.;
    \item \textit{Data analysis}: A.B., L.M.;
    \item \textit{Writing the manuscript}: M.B., A.B., L.M., D.G.P., F.G..
\end{itemize}



\begin{acks}
    This research has received funding from the \grantsponsor{823783}{European Union's Horizon 2020 FET Proactive} . 
    project ``WeNet - The Internet of us'', grant agreement No. \grantnum[https://doi.org/10.3030/823783]{823783}{823783}.  We deeply thank all the volunteers across the world for their participation in the study. We thank the anonymous reviewers for their valuable feedback. We acknowledge the use of ChatGPT and Grammarly as tools for grammar refinement.
    \unitn, \jlu, \num and \uc have participated to the work described in this paper as part of the DataScientia initiative (\url{https://datascientia.eu/}).
\end{acks}


\bibliographystyle{ACM-Reference-Format}
\bibliography{main}

\newpage
\appendix

\section{APPENDIX}
\subsection{Time Diaries}

This section provides a detailed outline of all the questions and answer options delivered through iLog. We report the diaries sent at fixed intervals throughout the study in the main text (\cref{tab:td_main}). Additionally, two time diaries were sent at the start and end of each day. The morning questions listed in \cref{tab:morning_questions} inquire about the sleep quality from the previous night and the participant’s expectations for the upcoming day. The evening questions, in \cref{tab:evening_questions}, ask participants to evaluate their day, mention any problems they encountered, describe how they addressed them, and report any challenges due to the COVID-19 pandemic.

To reduce the answering burden, participants could deactivate receiving questions for a fixed period by selecting one of the reasons listed in \cref{tab:break}.

The time diary questions were sent to participants every thirty minutes during the first two weeks and every hour in the final two weeks, asking about the activity being performed, location, social context, and mood. Additional questions were activated based on the activity selected in question \textit{A3} in \cref{tab:td_main} to gather more details. Specifically, when participants indicated they were traveling, they were asked about the purpose and mode of transportation (\textit{A3a1} and \textit{A3a2} in \cref{tab:td_sub}, respectively). Additional questions also prompted the type of food and drinks when participants were eating (\textit{A3c} in \cref{tab:td_sub}) and the type of sport when engaging in physical activities (\textit{A3b} in \cref{tab:td_sub}).

Every two hours, questions in \cref{tab:tb_snack} collect information about snacks and drinks outside the main meal periods.

\begin{table}[htb]
    \footnotesize
    \centering
    \caption{Morning questions sent at 8:00 AM.}
    \label{tab:morning_questions}
    \begin{tabularx}{0.97\textwidth}{p{8cm}p{8cm}}
    \toprule
    \textbf{A1. How would you rate your sleep quality last night?} &%
    \textbf{A2. How do you expect your day to be?}\\
    \midrule
    \begin{enumerate}[leftmargin=*]
    \item \vcenteredinclude{figures/emoji/emoji_1.png} very good
    \item \vcenteredinclude{figures/emoji/emoji_2.png} fairly good
    \item \vcenteredinclude{figures/emoji/emoji_3.png}
    \item \vcenteredinclude{figures/emoji/emoji_4.png} fairly bad
    \item \vcenteredinclude{figures/emoji/emoji_5.png} very bad
    \end{enumerate}
    &
    \begin{enumerate}[leftmargin=*]
    \item \vcenteredinclude{figures/emoji/emoji_1.png}
    \item \vcenteredinclude{figures/emoji/emoji_2.png}
    \item \vcenteredinclude{figures/emoji/emoji_3.png}
    \item \vcenteredinclude{figures/emoji/emoji_4.png}
    \item \vcenteredinclude{figures/emoji/emoji_5.png}
    \end{enumerate} \\
    \bottomrule
\end{tabularx}
    
    \caption{Evening questions sent at 10:00 PM.}
    \label{tab:evening_questions}
    \begin{tabularx}{0.97\textwidth}{XXX}
    \toprule
    \textbf{A7. How was your day?}&
    \textbf{A8. Did you have any problem at [college (weekdays)] today?}&
    \textbf{A9. What was the problem you had?} \\
    \midrule
    \begin{enumerate}[leftmargin=*]
    \item \vcenteredinclude{figures/emoji/emoji_1.png} 
        \item \vcenteredinclude{figures/emoji/emoji_2.png}
        \item \vcenteredinclude{figures/emoji/emoji_3.png}
        \item \vcenteredinclude{figures/emoji/emoji_4.png}
        \item \vcenteredinclude{figures/emoji/emoji_5.png}
    \end{enumerate}&
    \begin{enumerate}[leftmargin=*]
    \item Yes
    \item No
    \end{enumerate}&
    open-ended question\\
\end{tabularx}
\begin{tabularx}{0.95\textwidth}{XX}
    \toprule
    \textbf{A10.Were you able to solve the problem (alone or with help of someone)?}&
    \textbf{A11. Is there anything that you would have liked to do today that was not possible because of
the Covid-19 virus?} \\
    \midrule
    open-ended question & open-ended question \\
    \bottomrule
\end{tabularx}
\end{table}

\begin{table}[]
    \centering
    \caption{List of motivations to suspend question notifications for a fixed number of hours.}
    \label{tab:break}
    \begin{tabularx}{0.50\textwidth}{X}
        \toprule
        \textbf{Break options}\\
        \midrule
        \begin{enumerate}
        \setcounter{enumi}{35}
        \item Others
        \item I will participate in sports activities
        \item I have a work/study meeting
        \item I am at the cinema/theatre/hospital/church
        \item I am starting classes/lessons/lab
        \item I will go to sleep
        \end{enumerate} 
        \\
    \bottomrule
    \end{tabularx}
\end{table}

\begin{table}
    \footnotesize
    \centering
    \caption{In-depth questions that appear when certain options are selected in the question ``What are you doing?''}
    \label{tab:td_sub}
    \begin{tabularx}{0.97\textwidth}{XXXX}
    \toprule
    \textbf{A3a1. And you travel to/from or related to:}&
    \textbf{A3a2. How are you moving?}&
    \textbf{A3b. What kind of sports activity?}&
    \textbf{A3c. Select the main food \& drink you ate} (Multiple choices)\\
    \midrule
    \begin{enumerate}[leftmargin=*]
        \item study
        \item social life
        \item shopping and services
        \item other leisure
        \item work
        \item changing locality
        \item other or unspecified travel purpose
    \end{enumerate}&
    \begin{enumerate}[leftmargin=*]
        \item on foot
        \item by bike
        \item by bus/tram
        \item by metro/subway/underground
        \item by train
        \item by e-scooter
        \item by car
        \item by car as passenger
        \item by car sharing
        \item by moped, motorbike
        \item by moped, motorbike as passenger
        \item by motorboat
        \item by airplane
        \item by taxi/Uber
        \item other private transport modes
        \item other public transport modes
    \end{enumerate}&
    \begin{enumerate}[leftmargin=*]
        \item Walking, Trekking, and hiking
        \item Jogging and running
        \item Cycling, skiing, and skating
        \item Ball games
        \item Gymnastics and Fitness
        \item Water sports
        \item Other or unspecified sports or indoor activities
        \item Other or unspecified sports or outdoor activities
        \item Productive exercise (e.g., hunting, fishing, picking berries, mushrooms, or herbs)
    \end{enumerate}&
    \begin{enumerate}[leftmargin=*]
        \item Bread, steamed buns and/or breakfast cereals
        \item Rice, potatoes, beans, pasta, noodles, dumplings, etc.
        \item Vegetables
        \item Fruits
        \item Meat
        \item Fish
        \item Processed meat (ham, bacon, sausages)
        \item Dairy products (Plain or low-fat milk, yogurt, cheese)
        \item Soya-based food (milk, yogurt, tofu)
        \item Pastries and sweets
        \item Snack/sandwiches (chips\dots)
        \item Water
        \item Soda
        \item Coffee/tea or similar
        \item Others non-alcoholic drink
        \item Beer
        \item Wine
        \item Liquor
        \item Other alcoholic drink
        \item Other food
    \end{enumerate}
    \\
    \bottomrule
\end{tabularx}

\end{table}

\begin{table}
    \footnotesize
    \centering
    \caption{Additional questions related to food and drinks.}
    \label{tab:tb_snack}
    \begin{tabularx}{0.97\textwidth}{XX}
    \toprule
    \textbf{A6b. In the last two hours did you have any snacks or drinks (except breakfast, lunch, and dinner).} (administered at hours 02, 04, 06, 10, 12, 15, 17, 19, 22, 24) (Multiple choices) &
    \textbf{A6c. Select the food \& drink taken as snack. If you had more than one snack in the last two hours, only focus on the most recent one.} (Multiple choices)\\
    \midrule
    \begin{enumerate}[leftmargin=*]
        \item No
        \item Yes, between now and 30 minutes ago \textbf{(go to A6c)}
        \item Yes, between 0.5 and 1 hour ago \textbf{(go to A6c)}
        \item Yes, between 1 and 1.5 hours ago \textbf{(go to A6c)}
        \item Yes, between 1.5 and 2 hours ago \textbf{(go to A6c)}
    \end{enumerate}&
    \begin{enumerate}[leftmargin=*]
        \item Confectionery (Candy, Chocolate, etc)
        \item Cookies, cakes, and pastries
        \item Bars (Energy bar, etc.)
        \item Crackers/biscuits
        \item Seeds, nuts, grains, legumes
        \item Savory snacks (Chips, Tapas, Pizza, Nachos, Snack mix, deep frying)
        \item Sandwiches (Sandwich, Hamburgers, Hot dogs, Bagel)
        \item Frozen (Ice cream, Milkshake, etc.)
        \item Bread, steamed buns and/or breakfast cereals
        \item Rice, potatoes, beans, pasta, noodles,
        dumplings, etc.
        \item Vegetables
        \item Fruits
        \item Dairy products (milk, yogurt, cheese)
        \item Soya-based food (milk, yogurt, tofu)
        \item Meat
        \item Fish
        \item Processed meat (ham, bacon, sausages)
        \item Water
        \item Soda
        \item Coffee/tea or similar
        \item Others non-alcoholic drink
        \item Beer
        \item Wine
        \item Spirit
        \item Others alcoholic drink
        \item Other food
    \end{enumerate}\\
    \bottomrule
\end{tabularx}
\end{table}

\newpage 

\subsection{Sensors Description}\label{app2:sensors}

The smartphone sensors used by iLog are categorized as follows:
\begin{itemize}
    \item \textit{Hardware (HW)}, physical sensors of the device that detect and respond to physical environment, e.g., accelerometer, gyroscope and GPS;
    \item \textit{Software (SW)}, software component collecting events from the operating system and software, for instance, the Wifi the phone is connected to.
\end{itemize}
\cref{tab:sensor-list} reports the collection frequency of each sensor. The possible frequency values are:
\begin{itemize}
    \item \textit{on change} means that the value of the sensor is recorded only when the current value changes (along with a timestamp of when it happened);
    \item\textit{up to X samples per second} means that a maximum of X events are recorded for each second (the actual frequency in the data might slightly diverge from the reported one due to the device's inaccuracy);
    \item \textit{once every Y} means a new event is generated once every Y (also in this case the actual frequency approximates the reported one).
\end{itemize}
In \cref{tab:sensor}, we briefly describe each sensor and collected variable. The sensors are grouped by category, which corresponds to the distribution bundles (see \cref{sec:availability}).

\begin{table}[tb]
\centering
\caption{\label{tab:sensor-list} List sensors. The type column reports HW for hardware sensors, and SW for software sensors.}
\begin{tabular}{rclcc}
    \toprule
    \textbf{No} & \textbf{Type} & \textbf{Name} & \textbf{Frequency} \\
    \midrule
    1  & SW & Bluetooth Devices                 & Once every minute\\
    2  & SW & Cellular network info             & Once every minute\\
    3  & SW & WIFI Network Connected to         & On change\\
    4  & SW & WIFI Networks Available           & Once every minute\\
    5  & HW & Light                             & up to 10 samples per second \\
    6  & HW & Pressure                          & up to 10 samples per second\\
    7  & HW & Accelerometer                     & up to 10 samples per second \\ 
    8  & HW & Gyroscope                         & up to 10 samples per second  \\
    9 & SW & Movement Activity Label           & Once every 30 seconds\\
    10 & SW & Step Counter                      & up to 10 samples per second\\
    11 & SW & Step Detection                    & On change\\
    12 & HW & Location                          & Once every minute  \\
    13 & HW & Magnetic Field                    & up to 10 samples per second  \\
    14 & SW & Proximity                         & up to 10 samples per second\\
    15 & SW & Headset Status [ON/OFF]           & On change\\
    16 & SW & Music Playback & On change\\
    17 & SW & Notifications received            & On change\\
    18 & SW & Running Applications              & Once every 5 seconds\\
    19 & SW & Airplane Mode [ON/OFF]            & On change\\
    20 & SW & Battery Charge [ON/OFF]           & On change\\
    21 & SW & Battery Level                     & On change\\
    22 & SW & Doze Mode [ON/OFF]                & On change\\
    23 & SW & Ring mode [Silent/Normal]         & On change\\
    24 & SW & Touch event                       & On change\\
    25 & SW & Screen Status [ON/OFF]            & On change\\
    26 & SW & User Presence                     & On change\\
    \bottomrule
\end{tabular}
\end{table}

\begin{center}
\renewcommand{\arraystretch}{1.5}
\begin{longtable}{p{0.15\linewidth}|p{0.83\linewidth}}
    \caption[ilog sensors]{Description of the collected ilog sensors.} \label{tab:sensor} \\
    \toprule
    \textbf{Sensor} & \textbf{Description} \\
    \midrule
    \endfirsthead
    \multicolumn{2}{l}{\textbf{\textsc{Connectivity}}} \\
    Bluetooth & 
    List of discovered Bluetooth normal or low energy devices around the smartphone containing the following information:
    \begin{itemize}
        \item \textit{name}: user-friendly name of the remote device;
        \item \textit{address}: hardware MAC Address of the device;
        \item \textit{bondstate}: whether the remote device is connected;
        \item \textit{rssi}: Received Signal Strength Indicator;
        \item \textit{class code} and \textit{class tag}: Bluetooth class of the device (e.g., phone or computer), and the class describes the characteristics and capabilities of the device (e.g., audio and telephony).
    \end{itemize}\\
    Cellular Network &
    Information of the cellular network to which the phone is connected to:
    \begin{itemize}
        \item \textit{cellid}: identifier of the cell;
        \item \textit{dbm}: signal strength;
        \item \textit{type}: type of the cell, possible values are lte, wcdma, gsm and cdma.
    \end{itemize}\\
    WiFi Event & 
    Returns information related to the WIFI network to which the phone is connected; if connected, it also reports the WIFI network ID. Additional features are:
    \begin{itemize}
        \item \textit{ssid}: (Service Set Identifier) ID or unique identifier of a digital network (Wi-Fi or WLAN);
        \item \textit{bssid}: (Basic Service Set Identifier): sequence of characters that define a wireless computer network configured to communicate directly with each other;
        \item \textit{isconnected}: return whether the phone is connected to the WIFI.
    \end{itemize}\\
    WiFi Networks Event &
    Returns all WIFI networks detected by the smartphone. Additional features are:
    \begin{itemize}
        \item \textit{address}: is a unique identifier assigned to a network interface controller for use as a network address in communications within a network segment;
        \item \textit{capabilities}: list of capabilities supported by the network, e.g., WPA and WPS;
        \item \textit{frequency}: the WiFi frequency bands include 2.4 GHz and 5 GHz;
        \item \textit{name}: the name assigned to the WiFi network
        \item \textit{rssi}: Received Signal Strength Indicator is an estimated measure of signal strength that indicates how effectively a device can receive signals from any wireless access point or Wi-Fi router. It provides insight into the quality and reliability of the connection, often measured in decibels (dBm) to represent signal strength.
        The RSSI value range is between 0 and -100, where 0 signifies stronger and more stable connections.
    \end{itemize}\\
    \multicolumn{2}{l}{\textbf{\textsc{Motion}}} \\
    Accelerometer & 
    Measures the acceleration to which the phone is subjected and captures it as a 3D vector. The unit is $m/s^2$. \\
    Gyroscope &
    Measures the rotational forces to which the phone is subjected and it captures it as a 3D vector. The unit is $rad/s$. \\
    Activities &
    It reports the user's activity recognized by the Google Activity Recognition API. The recognized activities are \textit{in vehicle}, \textit{on bicycle}, \textit{on foot}, \textit{running}, \textit{still}, \textit{tilting}, \textit{walking} and \textit{unknown}. The sensor reports a confidence score between 0 and 100, which represents the likelihood that the user is performing the activity. \\
    Step Counter &
    It counts the total number of steps performed by the user (while carrying the phone) since the phone was powered on.\\
    Step detector & 
    An event is triggered each time the user takes a step. \\
    %
    %
    \multicolumn{2}{l}{\textbf{\textsc{Position}}} \\
    Location & It provides the geographic coordinates of the phone:
    \begin{itemize}
        \item \textit{latitude}: latitude in degrees;
        \item \textit{longitude}: longitude in degrees;
        \item \textit{altitude}: the altitude in meters;
        \item \textit{accuracy}: estimated horizontal accuracy radius in meters;
        \item \textit{speed}: current speed of the phone in $m/s$;
        \item \textit{provider}: the source of the coordinates, i.e., GPS (hardware sensor in the devices), network (based on the WiFi network  the phone is connected to) and passive (retrieve the location from other applications that already requested it);
        \item \textit{bearing}: horizontal direction of travel and it's the angle with respect to the north that is being faced.
    \end{itemize} \\
    Magnetic field &
    Reports the ambient magnetic field along the three sensor axes at the phone location. \\
    Proximity & Measures the distance between the user's head and the phone. Depending on the phone, it may be measured in centimeters (i.e., the absolute distance) or as labels (e.g., 'near', 'far'). \\
    \multicolumn{2}{l}{\textbf{\textsc{App usage}}} \\
    Headset status & Indicates whether the headphones are connected to the phone. \\
    Music Playback & Returns whether music is being played on the phone using the default music player from the operating system. Track information is not collected.\\
    Notifications & It generates an event every time the phone receives a notification and when it is dismissed by the user:
    \begin{itemize}
        \item \textit{identifier}: the unique identifier of notification within the application that generated it
        \item \textit{isclearable}: whether the notification can be canceled when the user clears the notifications;
        \item \textit{isongoing}: the notification refers to an event that is ongoing, e.g., a phone call;
        \item \textit{package}: package name of the application.
        \item \textit{status}: whether the notification is posted or dismissed.
    \end{itemize}\\
    Applications & Reports the name of the application (or application package) currently running in the foreground of the phone. \\
    \multicolumn{2}{l}{\textbf{\textsc{Device usage}}} \\
    Airplane Mode &  Returns whether the phone’s airplane mode is on or off. When off, all the connectivity features of the phone are turned off. Airplane mode also conserves battery life by reducing power-consuming background activities. It’s accessible through the quick settings menu on most devices. \\
    Battery Charge &
    Returns whether the phone is on charge:
    \begin{itemize}
        \item \textit{source}: type of power source connected to the device; possible values are USB, AC charge, wireless power source or unknown.
        \item \textit{status}: whether the device is being charged
    \end{itemize}\\
    Battery Monitoring Log & 
    Returns the phone's battery level:
    \begin{itemize}
        \item \textit{level}: current battery level between 0 and 100;
        \item \textit{scale}: maximum level of the battery represented as a value between 0 and 100.
    \end{itemize}\\
    Doze Mode & Returns whether the phone’s doze mode is on or off.
    Doze mode is a low-power state that a phone enters after a period of inactivity to conserve battery. In this mode, background processes and network access are restricted, allowing only essential tasks, such as high-priority notifications or alarms, to function periodically. This helps significantly reduce battery consumption while the device is idle.\\
    Ring Mode & Reports the current ring status of the phone. When set to \textit{normal}, the smartphone rings on incoming calls, messages and notifications by producing audible alerts. Other statuses are \textit{vibrate} (where the phone vibrates instead of ringing) or \textit{silent} (where all sounds are muted). Ring mode is often managed through the device's sound settings. \\
    Touch event & It generates an event each time the user touches the screen. \\
    Screen status & Returns whether the phone’s screen is on or off.\\
    User Presence & Detects when the user is present near the phone, for example, when the user unlocks the screen. \\
    \multicolumn{2}{l}{\textbf{\textsc{Enviroment}}} \\
    Light &  This component detects ambient light level around the phone, and it is measured in illuminance (lux). 
    This sensor helps the device adjust the screen brightness automatically, optimizing visibility while conserving battery life. For example, in bright conditions, the screen brightness increases for better readability, whereas in darker environments, it dims to reduce eye strain and save power. Light sensors are commonly located near the top of the device, often beside the front-facing camera.
    \\
    Pressure & It measures the ambient air pressure to which the phone is subjected in hPa or mbar. \\
    \bottomrule
\end{longtable}
\end{center}

\subsection{Dataset Files Structure}

The folder structure of the dataset is

\begin{lstlisting}[language=bash]
. dataset root
|- Site_Trento_ITA
   |- Diachronic-Interactions
      - timediaries.parquet
   |- Synchronic-Interactions
      |- matching.csv
      |- survey1.parquet
      |- survey2.parquet
      |- survey3.parquet
   |- Sensors
      |- App-usage
         |- application.parquet
         |- headsetplug.parquet
         |- music.parquet
         |- notification.parquet
      |- Connectivity
         |- bluetooth.parquet
         |- cellularnetwork.parquet
         |- wifi.parquet
         |- wifinetworks.parquet
      |- Device-usage
         |- airplanemode.parquet
         |- doze.parquet
         |- touch.parquet
         |- batterycharge.parquet
         |- ringmode.parquet
         |- userpresence.parquet
         |- batterymonitoringlog.parquet
         |- screen.parquet
      |- Environment
         |- light.parquet
         |- pressure.parquet
      |- Motion
         |- accelerometer.parquet
         |- gyroscope.parquet
         |- stepdetector.parquet
         |- activitiespertime.parquet
         |- stepcounter.parquet
      |- Position
         |- location_poi.parquet
         |- magneticfield.parquet
         |- location_rd.parquet
         |- proximity.parquet
|- Site_Copenhagen_DEN
   |- Diachronic-Interactions
      |-...
   |- Synchronic-Interactions
      |-...
   |- Sensors
      |-...
...
\end{lstlisting}
The file \texttt{matching.csv} maps the identifier of the participants in the sensor data with the identifier used in the surveys.

\end{document}